\documentclass[12pt]{article}
\usepackage{amsmath,amsthm,latexsym,amssymb,amsfonts,epsfig,psfrag}

\makeatletter \@addtoreset{equation}{section} \makeatother

\newtheorem{Lemma}{Lemma}[section]
\newtheorem{Corollary}{Corollary}[section]

\def\be{\begin{equation}}
\def\ee{\end{equation}}
\def\ba{\begin{eqnarray}}
\def\ea{\end{eqnarray}}

\def\Tau{\mathcal{T}}
\def\Nl{{\mathchoice
{\setbox0=\hbox{$\displaystyle\rm N$}\hbox{\hbox to0pt
{\kern0.4\wd0\vrule height0.9\ht0\hss}\box0}}
{\setbox0=\hbox{$\textstyle\rm N$}\hbox{\hbox to0pt
{\kern0.4\wd0\vrule height0.9\ht0\hss}\box0}}
{\setbox0=\hbox{$\scriptstyle\rm N$}\hbox{\hbox to0pt
{\kern0.4\wd0\vrule height0.9\ht0\hss}\box0}}
{\setbox0=\hbox{$\scriptscriptstyle\rm N$}\hbox{\hbox to0pt
{\kern0.4\wd0\vrule height0.9\ht0\hss}\box0}}}}
\def\Zl{{\mathchoice
{\setbox0=\hbox{$\displaystyle\rm Z$}\hbox{\hbox to0pt
{\kern0.4\wd0\vrule height0.9\ht0\hss}\box0}}
{\setbox0=\hbox{$\textstyle\rm Z$}\hbox{\hbox to0pt
{\kern0.4\wd0\vrule height0.9\ht0\hss}\box0}}
{\setbox0=\hbox{$\scriptstyle\rm Z$}\hbox{\hbox to0pt
{\kern0.4\wd0\vrule height0.9\ht0\hss}\box0}}
{\setbox0=\hbox{$\scriptscriptstyle\rm Z$}\hbox{\hbox to0pt
{\kern0.4\wd0\vrule height0.9\ht0\hss}\box0}}}}
\def\Ql{{\mathchoice
{\setbox0=\hbox{$\displaystyle\rm Q$}\hbox{\hbox to0pt
{\kern0.4\wd0\vrule height0.9\ht0\hss}\box0}}
{\setbox0=\hbox{$\textstyle\rm Q$}\hbox{\hbox to0pt
{\kern0.4\wd0\vrule height0.9\ht0\hss}\box0}}
{\setbox0=\hbox{$\scriptstyle\rm Q$}\hbox{\hbox to0pt
{\kern0.4\wd0\vrule height0.9\ht0\hss}\box0}}
{\setbox0=\hbox{$\scriptscriptstyle\rm Q$}\hbox{\hbox to0pt
{\kern0.4\wd0\vrule height0.9\ht0\hss}\box0}}}}
\def\Rl{{\mathchoice
{\setbox0=\hbox{$\displaystyle\rm R$}\hbox{\hbox to0pt
{\kern0.4\wd0\vrule height0.9\ht0\hss}\box0}}
{\setbox0=\hbox{$\textstyle\rm R$}\hbox{\hbox to0pt
{\kern0.4\wd0\vrule height0.9\ht0\hss}\box0}}
{\setbox0=\hbox{$\scriptstyle\rm R$}\hbox{\hbox to0pt
{\kern0.4\wd0\vrule height0.9\ht0\hss}\box0}}
{\setbox0=\hbox{$\scriptscriptstyle\rm R$}\hbox{\hbox to0pt
{\kern0.4\wd0\vrule height0.9\ht0\hss}\box0}}}}
\def\Cl{{\mathchoice
{\setbox0=\hbox{$\displaystyle\rm C$}\hbox{\hbox to0pt
{\kern0.4\wd0\vrule height0.9\ht0\hss}\box0}}
{\setbox0=\hbox{$\textstyle\rm C$}\hbox{\hbox to0pt
{\kern0.4\wd0\vrule height0.9\ht0\hss}\box0}}
{\setbox0=\hbox{$\scriptstyle\rm C$}\hbox{\hbox to0pt
{\kern0.4\wd0\vrule height0.9\ht0\hss}\box0}}
{\setbox0=\hbox{$\scriptscriptstyle\rm C$}\hbox{\hbox to0pt
{\kern0.4\wd0\vrule height0.9\ht0\hss}\box0}}}}
\def\Hl{{\mathchoice
{\setbox0=\hbox{$\displaystyle\rm H$}\hbox{\hbox to0pt
{\kern0.4\wd0\vrule height0.9\ht0\hss}\box0}}
{\setbox0=\hbox{$\textstyle\rm H$}\hbox{\hbox to0pt
{\kern0.4\wd0\vrule height0.9\ht0\hss}\box0}}
{\setbox0=\hbox{$\scriptstyle\rm H$}\hbox{\hbox to0pt
{\kern0.4\wd0\vrule height0.9\ht0\hss}\box0}}
{\setbox0=\hbox{$\scriptscriptstyle\rm H$}\hbox{\hbox to0pt
{\kern0.4\wd0\vrule height0.9\ht0\hss}\box0}}}}
\def\Ol{{\mathchoice
{\setbox0=\hbox{$\displaystyle\rm O$}\hbox{\hbox to0pt
{\kern0.4\wd0\vrule height0.9\ht0\hss}\box0}}
{\setbox0=\hbox{$\textstyle\rm O$}\hbox{\hbox to0pt
{\kern0.4\wd0\vrule height0.9\ht0\hss}\box0}}
{\setbox0=\hbox{$\scriptstyle\rm O$}\hbox{\hbox to0pt
{\kern0.4\wd0\vrule height0.9\ht0\hss}\box0}}
{\setbox0=\hbox{$\scriptscriptstyle\rm O$}\hbox{\hbox to0pt
{\kern0.4\wd0\vrule height0.9\ht0\hss}\box0}}}}
\title{Improved and Perfect Actions in Discrete Gravity}
\author{Benjamin Bahr$^1$ and Bianca Dittrich$^2$\\
\small $^1$ DAMTP, University of Cambridge,\\
\small  Wilberforce Road, Cambridge CB3 0WA, UK \\
\small   $^2$ MPI f. Gravitational Physics, Albert Einstein Institute,\\
 \small Am M\"uhlenberg 1, D-14476 Potsdam, Germany\\
\small and\\
\small Institute for Theoretical Physics, Utrecht University,\\
\small  Leuvenlaan 4, NL-3584 CE Utrecht, The Netherlands }

\begin{document}

\maketitle

\begin{abstract}
We consider the notion of improved and perfect actions within Regge
calculus. These actions are constructed in such a way that they -
although being defined on a triangulation -  reproduce the continuum
dynamics exactly, and therefore capture the gauge symmetries of
General Relativity. We construct the perfect action in three
dimensions with cosmological constant, and in four dimensions for
one simplex. We conclude with a discussion about Regge Calculus with
curved simplices, which arises naturally in this context.
\end{abstract}


\section{Introduction}

In General Relativity the notion of diffeomorphism invariance,
resulting from Einstein's covariance principle, is of ultimate
importance \cite{WALDGR}. In particular, its correct implementation
on the quantum level is a challenging task for every candidate
quantum gravity theory.

Prior to quantizing a classical field theory, it is usually first
discretized, since discrete systems generically have finitely many
degrees of freedom. These are usually easier to quantize than the
infinitely many degrees of freedom of field theories. A natural
discretization of General Relativity is achieved by Regge calculus,
where the smooth space-time is replaced by a simplicial complex, and
the metric information is contained in the edge lengths and deficit
angles around the hinges \cite{REGGE, QREGGE}. A similar
discretization is used within the Spin Foam quantization approach,
where the variables of the first order Plebanski formulation of GR
are discretized on Regge triangulations, prior to quantization
\cite{SF}.

It is an important question what happens with the diffeomorphism
invariance of General Relativity in these discretized gravity
theories (see \cite{BIANCASYMM} and references therein).\\[5pt]

Discretizing a theory often breaks symmetries, such as in QCD, where
the introduction of a lattice breaks i.e. rotation invariance.
Another example are reparametrization invariant one-dimensional
systems, where the discretization scheme generically breaks the
reparametrization invariance \cite{marsden}. The latter example
resembles the situation in GR in many ways \cite{OURS}.

In a canonical formulation the problem becomes even more apparent,
where the symmetries turn into constraints, and it is notoriously
difficult to implement them correctly in the discretized quantum
theory (see also \cite{GP} for a discussion). Even in quantum
gravity theories like LQG, which are inherently set out to capture
the full continuum physics, the discretized nature of the
constituents, i.e. the graphs, make the implementation of the
constraint algebra rather non-trivial \cite{BIANCASYMM, QSD8}.\\[5pt]

In general, breaking of symmetries is, however, not ultimately tied
to the discretization, but rather the approximation involved, i.e.
by replacing spatial derivatives with differential quotients between
neighboring lattice points.

For instance in lattice QCD one ideally would want to construct a
Lagrangian which, although describing a theory on the lattice, still
encodes the symmetries of the continuum theory \cite{LQCDSCRIPT,
LQCD1}. A lattice action which reproduces the same dynamics as the
continuum theory and therefore also reflects the symmetries of the
continuum limit is termed \emph{perfect action} in that context.
That perfect actions exist for asymptotically free theories follows
from Wilson's theory of renormalization group flow \cite{RGWILSON}.
Although for actual problems at hand the perfect actions are very
hard to compute, the (numerical) computation of \emph{improved
actions}, i.e. actions that capture the continuum symmetries much
better than the actual na\"ive lattice discretization, is an
important task. These actions are widely sought for in order to
suppress lattice artifacts in numerical calculations \cite{LQCD2}.\\[5pt]

In this article we investigate the question of how improved and
perfect actions within the context of discretizations of General
relativity, in particular Regge calculus, can be constructed. We
will start with reviewing one-dimensional reparametrization
invariant systems and their discretization in chapter
\ref{Ch:DiscretizedActionsIn1D}. These systems exhibit a gauge
symmetry which mimics diffeomorphism symmetry of GR in many
respects. This symmetry is broken in the na\"ive discretization of
those systems, and we will have a look at how one can construct
improved and perfect actions for them. In particular we will see how
the perfect actions restore the gauge invariance of the continuum
limit within the discretized setting. Part of this chapter will
follow \cite{marsden}.

In chapters \ref{Ch:ReggeCalculus} and
\ref{Ch:RefinementOfTheReggeAction} we will focus on Regge calculus
with a cosmological constant in three and four dimensions. Whereas
Regge calculus in $3D$ with $\Lambda=0$ exhibits a well-known vertex
displacement symmetry which is a result of the discrete Bianchi
identities \cite{HERBIEBIANCHI, FREILOU}, this symmetry is broken
for $\Lambda\neq 0$. We show how to construct improved actions in
this case and analytically compute the perfect action, which regains
the vertex displacement symmetry and hence reflects the dynamics and
the symmetry of the continuum, albeit formulated on a Regge
triangulation.

We also formulate improved actions for Regge calculus in $4D$, and
investigate some properties of its continuum limit, i.e. the
corresponding perfect action. In particular we are able to show that
the perfect action from the Regge action, and the one obtained by
using simplices of constant curvature instead of internally flat
ones, coincide. In the language of renormalization group flow this
demonstrates that the two actions one started with lie in the same
universality class.

We will in particular comment about the conclusions one can draw
from these findings for the corresponding quantum theories.

\section{Discretized actions in 1D}\label{Ch:DiscretizedActionsIn1D}

In this section we will discuss theories arising from
discretizations of systems with one--dimensional reparametrization
invariance, that is invariance under redefinitions of the time
variable. As we will see under discretization the exact
reparametrization invariance is typically lost similar to the
diffeomorphism invariance in the Regge action. However for the
examples we consider in this section there is a procedure to obtain
a discrete action with an exact reparametrization invariance. This
procedure resembles the ``blocking from the continuum'' construction
in lattice QCD, where a lattice action is constructed which has the
exact symmetries of the continuum action \cite{LQCD2}. In some parts
of our discussion we will follow \cite{marsden}.\\[5pt]

We start from a regular Lagrangian $L(q,\dot q)$ where $q$ denotes
the configuration variable. We assume that the dynamics determined
by $L$ leads to a unique solution $q(t)$ for given boundary values
$q(t_i)$, $q(t_f)$ if $t_i$ and $t_f$ are sufficiently close
together.

From this we construct a reparametrization invariant action by
adding the time variable $t$ to the configuration variables and use
$s$ as an (auxiliary) evolution parameter instead. If we define

\begin{eqnarray}\label{Gl:ReparametrizationInvariantLagrangianIn1D}
\tilde L(t,q,t',q')\;:=\;L\left(q, \frac{{q}'}{t'}\right)\, t',
\end{eqnarray}

\noindent where a prime denotes differentiation with respect to $s$,
then it is then straightforward to verify that

\begin{eqnarray}\label{Gl:ReparametrizationInvariantActionIn1D}
S\;=\;\int_{s_i}^{s_f} \tilde L(t,q,t',q')\, ds
\end{eqnarray}

\noindent is indeed invariant under reparametrizations $\tilde s=
f(s)$ of the evolution parameter and the induced change $\tilde
t(\tilde s)=t(f^{-1}(\tilde s)), \tilde q(\tilde s)=q(f^{-1}(\tilde
s))$ of the evolution paths.\\

The Euler-Lagrange equations for
(\ref{Gl:ReparametrizationInvariantActionIn1D}) for the variables
$t,q$ are given by
\begin{eqnarray}\label{Gl:EulerLagrangeForInvariantSystem1D_1}
\left(\frac{\partial \tilde L}{\partial
q}\,-\,\frac{d}{ds}\frac{\partial \tilde L}{\partial
q'}\right)\;&=&\;\frac{\partial L}{\partial
q}\,t'\,-\,\frac{d}{ds}\frac{\partial L}{\partial \dot q}\;=\;
\left(\frac{\partial L}{\partial q}\,-\,\frac{d}{d
t}\frac{\partial L}{\partial \dot q}\right)t'\\[5pt]\label{Gl:EulerLagrangeForInvariantSystem1D_2}
\left(\frac{\partial \tilde L}{\partial
t}\,-\,\frac{d}{ds}\frac{\partial \tilde L}{\partial
t'}\right)\;&=&\;-\frac{d}{ds}L\;+\;\frac{\partial L}{\partial
q}\,q'\,+\,\frac{\partial L}{\partial \dot
q}\,\frac{d}{ds}\frac{dq'}{dt'}\;=\;\left(-\frac{dL}{dt}\,+\,\frac{\partial
L}{\partial q}\dot q\,+\,\frac{\partial L}{\partial \dot q}\ddot
q\right)t'
\end{eqnarray}

\noindent where $\frac{\partial L}{\partial q}$ and $\frac{\partial
L}{\partial \dot q}$ denote the derivative of $L$ w.r.t. its first
and its second entry respectively. Note that
(\ref{Gl:EulerLagrangeForInvariantSystem1D_1}) is equivalent to the
Euler-Lagrange equations for $L$, and
(\ref{Gl:EulerLagrangeForInvariantSystem1D_2}) is satisfied
identically due to the chain rule. So this is just a reformulation
of the dynamics determined by $L$ via introduction of a gauge degree
of freedom. The non-uniqueness of the solutions $t(s),\,q(s)$
directly corresponds to the reparametrization independence of the
dynamical system defined by the Lagrangian $\tilde L$.\\[5pt]

A na\"{i}ve discretization of the action
(\ref{Gl:ReparametrizationInvariantActionIn1D}) is given by
\begin{eqnarray}\label{Gl:DiscretizationOfReparametrizationInvariantActionIn1D}
S_d\;=\;\sum_{n=0}^{N-1} (t_{n+1}-t_n)  L_n
\end{eqnarray}

\noindent with

\begin{eqnarray}\label{Gl:DiscretizationOfReparametrizationInvariantLagrangeFunctionIn1D}
L_n\;:=\;L\left( q_n\,,\,\,
\frac{q_{n+1}-q_n}{t_{n+1}-t_{n}}\right).
\end{eqnarray}

\noindent The dynamics of this discretized system is obtained by
looking for stationary variations of
(\ref{Gl:DiscretizationOfReparametrizationInvariantActionIn1D})
w.r.t the $t_n,q_n$. The equations of motion are

\begin{eqnarray}\label{Gl:DiscreteVariationWRTQ}
0=\frac{\partial S_d}{\partial
q_n}\;&=&\;\partial_qL_n(t_{n+1}-t_n)\,+\,\partial_{\dot q}L_{n-1}-\partial_{\dot q}L_n\\[5pt]\label{Gl:DiscreteVariationWRTT}
0=\frac{\partial S_d}{\partial
t_n}\;&=&\;L_{n-1}-L_n\,+\,\partial_{\dot q}L_n\frac{q_{n+1}-q_n}{t_{n+1}-t_n}\\[5pt]\nonumber
&&\,-\,\partial_{\dot q}L_{n-1}\frac{q_{n}-q_{n-1}}{t_{n}-t_{n-1}}
\end{eqnarray}

\noindent where $\partial_{q}L_n$ denotes the derivative
$\frac{\partial L}{\partial q}$ evaluated at $q=q_n,\,\dot
q_n=\frac{q_{n+1}-q_n}{t_{n+1}-t_n}$. Similarly $\partial_{\dot
q}L_n$ is the derivative of $L$ w.r.t. its second entry evaluated at
$(q_n,\dot q_n)$. With the product rule
$A_{n+1}B_{n+1}-A_nb_n=A_{n+1}(B_{n+1}-B_n)+A_n(B_{n+1}-B_n)$,
equation (\ref{Gl:DiscreteVariationWRTT}) for the $t_n$ can, using
(\ref{Gl:DiscreteVariationWRTQ}), be rewritten as

\begin{eqnarray}\nonumber
0\;&=&\;-\frac{L_n-L_{n-1}}{t_{n+1}-t_n}\,+\,\partial_{q}L_n\frac{q_{n+1}-q_n}{t_{n+1}-t_n}\\[5pt]\label{Gl:DiscreteVersionOfEnergyConservation}
\,&+&\,\partial_{\dot
q}L_{n-1}\frac{1}{t_{n+1}-t_n}\left(\frac{q_{n+1}-q_n}{t_{n+1}-t_n}-\frac{q_{n}-q_{n-1}}{t_{n}-t_{n-1}}\right).
\end{eqnarray}

\noindent In the continuum limit
(\ref{Gl:DiscreteVersionOfEnergyConservation}) converges to
\begin{eqnarray} -\frac{dL}{dt}\,+\,\frac{\partial L}{\partial
q}\dot q\,+\,\frac{\partial L}{\partial \dot q}\ddot q
\end{eqnarray}

\noindent which vanishes identically, and is equivalent to
(\ref{Gl:EulerLagrangeForInvariantSystem1D_2}). In the discrete case
however, the equations of motion
(\ref{Gl:DiscreteVersionOfEnergyConservation}) for the $t_n$ do not
vanish in general. So the equations for the $t_n$ are nontrivial,
and have to be solved along with the $q_n$. Since the equations
(\ref{Gl:DiscreteVersionOfEnergyConservation}) only couple $t_n$ at
most two steps apart from each other, the discrete system is of
second order and generically imposing boundary values
$t_0,q_0,t_N,q_N$ uniquely determines a solution. As a consequence,
the discrete system defined by the action
(\ref{Gl:DiscretizationOfReparametrizationInvariantActionIn1D}) does
not capture the reparametrization invariance of the continuum
dynamics defined by (\ref{Gl:ReparametrizationInvariantActionIn1D}).
One can show that this is directly linked to the failure of energy
conservation within the the discrete system \cite{marsden, OURS,
GP}.

The loss of reparametrization invariance is, however, not ultimately
tied to the discretization itself, but rather to the approximation
(\ref{Gl:DiscretizationOfReparametrizationInvariantLagrangeFunctionIn1D}).
If, however, one can find a discrete action that exactly reproduces
the continuum dynamics, one can regain the reparametrization
freedom. Such actions are termed \emph{perfect actions} e.g. in
lattice gauge theory\footnote{Where, however, the broken symmetry in
question is usually global Poincar\'{e} symmetry, and not gauge
symmetries.}. In the following chapter we will show how to construct
a perfect action for the $1D$ systems discussed above, in order to
restore reparametrization invariance.

\subsection{Regaining reparametrization
invariance}\label{Ch:PerfectActionsIn1D}

For the type of discretized actions we discussed so far one can
always define a discrete action which displays exact
reparametrization invariance. This so-called perfect action reflects
the gauge freedom of the continuous system, which results in a
non-uniqueness of the solution $\{t_n,q_n\}$. The idea is that the
discrete system should exactly reproduce the dynamics of the
continuous system, determined by the continuum Lagrange function
$L(q,\dot q)$.

We define the perfect action as follows: For $t_n,q_n$, and for each
$n=0,\ldots N-1$ solve the continuum Euler-Lagrange equations for
$t^{(n)}(s),q^{(n)}(s), \;s\in[0,1]$ with boundary values

\begin{eqnarray}\label{Gl:EOMForHamiltonJacobi}
\begin{array}{rcl}t^{(n)}(0)&=&t_n\;\qquad  q^{(n)}(0)=q_n\\[5pt]
t^{(n)}(1)&=&t_{n+1}\;\qquad q^{(n)}(1)=q_{n+1}.\end{array}
\end{eqnarray}

\noindent Denote the value of the action $S$ on that solution, which
is nothing but the Hamilton-Jacobi functional, by $S^{(n)}_{HJ}$ and
define

\begin{eqnarray} \label{Gl:ExactDiscreteActionIn1D}
S_e\;&:=&\;\sum_{n=0}^{N-1}
S^{(n)}_{HJ}(t_n,q_n,t_{n+1},q_{n+1})\\[5pt]\nonumber
\;&=&\;\sum_{n=0}^{N-1}\int_{0}^{1}ds\;\tilde
L\left(t^{(n)}(s),q^{(n)}(s), t^{(n)'}(s), q^{(n)'}(s)\right)
\end{eqnarray}

\noindent where $t^{(n)'}$ and $q^{(n)'}$ denote the derivatives of
$t^{(n)}$ and $q^{(n)}$ w.r.t. the curve parameter $s$,
respectively, and $\tilde L$ is given by
(\ref{Gl:ReparametrizationInvariantLagrangianIn1D}).

The discrete action $S_e$ defined in
(\ref{Gl:ExactDiscreteActionIn1D}) is exactly reparametrization
invariant, as the following theorem shows.\\

\noindent {\bf Theorem:} For each solution $\{t_n,q_n\}$ of the
equations of motion determined by the action
(\ref{Gl:ExactDiscreteActionIn1D}) and each sequence $\{s_n\}$ there
is a solution $t(s)$, $q(s)$ of the equations of motion
(\ref{Gl:EulerLagrangeForInvariantSystem1D_1}),
(\ref{Gl:EulerLagrangeForInvariantSystem1D_2}) with
$t(s_n)=t_n,\,q(s_n)=q_n$. Furthermore, for every such solution
$t(s),\,q(s)$ and each $s_0<s_1\ldots<s_N$, $\{t(s_n),q(s_n)\}$ is a
solution to the equations of motion determined by
(\ref{Gl:ExactDiscreteActionIn1D}).\\

\noindent\textbf{Proof:} A detailed proof of this can be found in
\cite{marsden}. \\[5pt]

Since the continuous system with the Lagrangian $\tilde L$ is
reparametrization invariant, the solutions of
(\ref{Gl:EulerLagrangeForInvariantSystem1D_1}),
(\ref{Gl:EulerLagrangeForInvariantSystem1D_2}) are highly
non-unique. Therefore, also the boundary value problem for the
action (\ref{Gl:ExactDiscreteActionIn1D}) has a vast amount of
different solutions for the same boundary conditions. This
non-uniqueness directly corresponds to the reparametrization
invariance of the action
(\ref{Gl:ReparametrizationInvariantActionIn1D}), and hence the
discrete action $S_e$ exactly captures this invariance. In
particular, the $t_n,\,q_n$ are underdetermined. Given the
uniqueness of solutions to the dynamics determined by the
deparametrized system with Lagrangian $L(q,\dot q)$ - the $q_n$ are
uniquely determined by the $t_n$, which by themselves can be chosen
arbitrarily\footnote{As long as $t_n<t_{n+1}$ for all $n$, i.e. the
$t_n$ are a growing sequence.}. It follows that there is one gauge
degree of freedom per vertex. Note that the $q_n(t_n)$ are Dirac
observables in the sense of \cite{ROVELLI1}.\\[5pt]

We have seen that the discrete action
(\ref{Gl:ExactDiscreteActionIn1D}) exactly mimics the continuum
dynamics of the system and therefore exhibits exact
reparametrization invariance, unlike the system defined by the
na\"ive discretization
(\ref{Gl:DiscretizationOfReparametrizationInvariantActionIn1D}).
Note that, as the discretization becomes very fine, one can expect
the system to be approximately reparametrization invariant in the
sense of \cite{OURS}. The Hessian of $S_d$ at the solution will
contain a large number of Eigenvalues\footnote{Namely one per
(inner) vertex.} approaching zero in the continuum limit, when
reparametrization invariance is restored.\\[5pt]

\subsection{Improving the discrete action $S_d$}

The perfect action $S_e$ contains the Hamilton-Jacobi functional of
the system defined by the Lagrangian $L$, which might in general be
hard to compute, or even unknown. In the following we present a
procedure to construct sequences of \emph{improved} actions, which
converge to the perfect action, and which satisfy the constraints in
an approximate way \cite{OURS}.\\[5pt]

In order to improve the action $S_d$, which is a na\"ive
discretization of the action
(\ref{Gl:ReparametrizationInvariantActionIn1D}) on the discretized
interval $\{t_n\}$, one needs to refine the interval by $t_n=\tilde
t_{nM}<\tilde t_{nM+1}<\ldots<\tilde t_{nM+(M-1)}<\tilde
t_{(n+1)M}=t_{n+1}$. Fix $\{t_n,q_n\}$, and for each interval
$[t_n,t_{n+1}]$ solve the discrete equations of motion or the
$\tilde t_k,\tilde q_k$, given by the na\"ive discretization of the
action $S$, i.e. find an extremum of the action

\begin{eqnarray}\label{Gl:ImprovedActionIn1D}
S_d^{(n)}\;=\;\sum_{k=Mn}^{Mn+M-1}L\left(\tilde q_k,\frac{\tilde
q_{k+1}-\tilde q_k}{\tilde t_{k+1}-\tilde t_k}\right)(\tilde
t_{k+1}-\tilde t_k)
\end{eqnarray}

\noindent with the boundary conditions

\begin{eqnarray*}
\tilde t_{Mn}&=&t_n,\qquad \tilde q_{Mn}=q_n\\[5pt]
\tilde t_{M(n+1)}&=&t_{n+1},\qquad \tilde q_{M(n+1)}=q_{n+1}.
\end{eqnarray*}

\noindent Denote the value of $S_d^{(n)}$ on the solution by
$S_*^{(n)}$. Then the action

\begin{eqnarray}\label{Gl:ImprovedActionFor1D}
S_*\;:=\;\sum_{n=0}^{N-1}S_*^{(n)}
\end{eqnarray}

\noindent is clearly a function of the chosen $t_n,q_n$. It is more
complicated than the na\"ive discretization
(\ref{Gl:DiscretizationOfReparametrizationInvariantActionIn1D}).

Since for very fine subdivision the $\tilde t_k,\,\tilde q_k$
converge to a solution $t(s),\,q(s)$ of the continuum dynamics given
by $\tilde L$, it is easy to see that - in the limit of very fine
discretization $\tilde t_k$ each of the contributions $S_*^{(n)}$
converges to its continuous counterpart, i.e.

\begin{eqnarray}
\lim_{M\to\infty}S_*^{(n)}\;=\;S_{HJ}^{(n)}(t_n,q_n,t_{n+1},q_{n+1})\;=\;\int_{s_n}^{s_{n+1}}ds\;L(t(s),q(s)).
\end{eqnarray}

\noindent Therefore $S_*$ converges to the exact discrete action
(\ref{Gl:ExactDiscreteActionIn1D}).\\[5pt]

The na\"ively discretized action
(\ref{Gl:DiscretizationOfReparametrizationInvariantActionIn1D})
approximates the exact discrete action
(\ref{Gl:ExactDiscreteActionIn1D}) by replacing, for each interval
$[s_n,s_{n+1}]$, the integral over the Lagrangian by a Riemann sum
involving only two points. The improvement within the action $S_*$
lies in the fact that the Riemann sum used to approximate the
integral relies on many more intermediate points, therefore
delivering a better approximation.\\[5pt]

In order to compute the improved actions, only the solutions to the
na\"ively discretized action $S_d$ for a refined discretization is
involved, making the computation possibly more feasible, if the
continuum system is not at hand, or too difficult to solve.
Furthermore, the improved action $S_*$ can can be made an
arbitrarily good approximation to the exact discrete action $S_e$,
by using a very fine discretization, or by iterating the process,
i.e. computing $S_*,\,(S_*)_*,\,((S_*)_*)_*,\ldots$, which lead to
the same limit $S_e$. It can therefore be used to compute $S_e$
recursively, which can therefore be seen as the ``perfect limit'' of
the $S_*$. We will use this strategy in order to investigate the
perfect action in Regge gravity later on.

Note that although $S_*$ still does not retain the full
reparametrization invariance of $S_e$, it is closer to it than the
na\"ively discretized action $S_d$, in the sense that the
constraints are satisfied to a greater accuracy\footnote{See
\cite{OURS} for details on approximate constraints.}.\\[5pt]


\section{Regge Calculus}\label{Ch:ReggeCalculus}

In the previous system we have seen that one-dimensional
reparametrization invariant systems usually lose that invariance
after discretization. This is also true for higher-dimensional field
theories: Classical GR, as a theory of metrics on a differential
manifold, is reparametrization invariant, due to the principle of
covariance \cite{WALDGR}. It is this invariance which makes it very
difficult to compute, or interpret the physics of solutions. It also
is connected to many obstacles for quantizing the theory
\cite{BIANCASYMM, GP, ROVELLI, INTRO}.

Regge Calculus provides a discretization of GR, by triangulating the
manifold, and replacing curvature expressions with deficit angles
around $2$-codimensional subsimplices \cite{REGGE}. Just as the
discretization of one-dimensional systems replaces the search for
smooth solutions to the equations by piecewise linear ones, Regge
calculus replaces smooth curved metrics by piecewise linear flat
ones.

\subsection{Continuous preliminaries}\label{Ch:ContinuousPreliminaries}

The Einstein-Hilbert action in $D$ dimensions with cosmological
constant $\Lambda$ is given by
\begin{eqnarray*}
S_{EH}\;=\;\frac{1}{8\pi}\int_{\mathcal{M}}d^Dx\,\sqrt{|g|}\;\,\left(\Lambda\,-\,\frac{1}{2}R\right)
\end{eqnarray*}

\noindent leading to the equations of motion
\begin{eqnarray}\label{Gl:EinsteinEquationsWithLambda}
8\pi\frac{\partial S_{EH}}{\partial
g^{\mu\nu}}\;=\;R_{\mu\nu}-\frac{1}{2}g_{\mu\nu}R\,+\,\Lambda
g_{\mu\nu}\;=\;0
\end{eqnarray}

\noindent There is a special solution to
(\ref{Gl:EinsteinEquationsWithLambda}), which will be very important
later on. The Riemann tensor of a space of constant (sectional)
curvature $\kappa$ has the property
\begin{eqnarray}\label{Gl:RiemannOfConstantCurvature}
R_{\mu\nu\sigma\rho}\;=\,\kappa\big(g_{\mu\rho}g_{\nu\sigma}\,-\,g_{\mu\sigma}g_{\nu\rho}\big)
\end{eqnarray}

\noindent leading to $R_{\mu\nu}=\kappa(D-1)g_{\mu\nu}$ and
$R=\kappa D (D-1)$. Therefore the metric satisfying
(\ref{Gl:RiemannOfConstantCurvature}) satisfies the equations
(\ref{Gl:EinsteinEquationsWithLambda}) for
\begin{eqnarray}\label{Gl:LambdaAndKappaRelation}
\Lambda\;=\;\frac{(D-1)(D-2)}{2}\kappa.
\end{eqnarray}

\subsection{Discrete action}

In Regge calculus, the smooth manifold $\mathcal{M}$ is replaced by
a triangulated manifold $\Tau$, the $D$-simplices of which are
internally flat \cite{REGGE}. The Riemann curvature in this now
arises as nontrivial parallel transport resulting from the
nontrivial way of gluing the simplices together. The curvature is
therefore naturally associated to the $D-2$ simplices $H$ (also
called ``hinges'') in the complex.

The Ricci scalar for such a manifold is  $2$ times the deficit angle
at a $D-2$ simplex. If the triangulation $\Tau$ has a boundary
$\partial \Tau$ (which is a triangulated $D-1$-dimensional
manifold), then the action has a contribution from the extrinsic
curvature in the boundary and the Ricci curvature in the bulk
$\Tau^\circ:=\Tau\backslash\partial \Tau$, and reads\footnote{Up to
a factor of $8\pi$, which we ignore from now on.}
\cite{HARSOR-BOUNDARY}

\begin{eqnarray}\label{Gl:ReggeActionFlat}
S_{\Tau}\;=\;\sum_{h\in \Tau^\circ}
F_{h}\epsilon_h\,-\,\Lambda\sum_{\sigma\subset
\Tau^\circ}V_\sigma\;+\;\sum_{h\in\partial \Tau}F_h\psi_h
\end{eqnarray}

\noindent The sum goes over all $(D-2)$-simplices $h$ in the bulk
and the boundary separately. The associated angles are
\begin{eqnarray}\label{Gl:FlatDeficitAngle}
\epsilon_h\;&=&\;2\pi\,-\,\sum_{\sigma\supset
h}\theta_h^\sigma\qquad \text{for }h\in
\Tau^\circ\\[5pt]\label{Gl:FlatExteriorAngle}
\psi_h\;&=&\;\pi\,-\,\sum_{\sigma\supset h}\theta_h^\sigma\qquad
\text{for }h\in \partial \Tau
\end{eqnarray}

\noindent where the $\theta_h^\sigma$ is the interior dihedral angle
in the $D$-simplex $\sigma$ associated to the $D-2$-subsimplex
$h\subset\sigma$. For these angles within a flat simplex $\sigma$
the so-called Schlaefli-identity reads
\begin{eqnarray}\label{Gl:SchlaefliFlat}
\sum_{h\subset \sigma}F_h\frac{\partial \theta_h^{\sigma}}{\partial
l_e}\;=\;0\qquad\text{for all $1$-simplices (``edges'')
}e\subset\sigma
\end{eqnarray}

\noindent The dynamical variables are taken to be the lengths $l_e$
of the edges $e\in\mathcal{T}^\circ$ in the bulk. For those edges
the equations of motion can be computed with
(\ref{Gl:SchlaefliFlat}) to be:
\begin{eqnarray}\label{Gl:ReggeEquationsFlat}
\sum_{h\supset e}\frac{\partial F_h}{\partial
l_e}\epsilon_h\;-\;\Lambda\sum_{\sigma\supset e}\frac{\partial
V_\sigma}{\partial l_e}\;=\;0.
\end{eqnarray}

\noindent Instead of piecewise linear flat simplices, one can build
up the triangulation with simplices of constant (sectional)
curvature $\kappa$ (see appendix \ref{Ch:AppendixCurvedSimplices}).
The Regge action for such a triangulation $\Tau$ with cosmological
constant $\Lambda$ is a sum of the overall curvature of the
manifold, having a contribution from the deficit angles at the $D-2$
dimensional subsimplices, the constant curvature of the tetrahedra,
and the term with the cosmological constant. For $\Lambda$ and
$\kappa$ having the relation (\ref{Gl:LambdaAndKappaRelation}), this
leads to
\begin{eqnarray}\label{Gl:ReggeActionWithCurvedSimplices}
S_{\Tau}^{(\kappa)}\;=\;\sum_{h\subset \Tau^\circ}
F^{(\kappa)}_h\epsilon_h^{(\kappa)}\,+\,(D-1)\kappa\sum_{\sigma\subset
\Tau^\circ}V_\sigma^{(\kappa)}\;+\;\sum_{h\in\partial
\Tau}F^{(\kappa)}_h\psi_h^{(\kappa)}.
\end{eqnarray}

\noindent where $F_h^{(\kappa)}$ denotes the $D-2$-dimensional
volume of the $D-2$ simplex $h\subset\sigma$. Furthermore
$\epsilon^{(\kappa)}_h$ and $\psi^{(\kappa)}_h$ denote deficit angle
and exterior angle in the curved simplices analogously to
(\ref{Gl:FlatDeficitAngle}), (\ref{Gl:FlatExteriorAngle}).

The Schlaefli identity (\ref{Gl:SchlaefliCurved}) for curved
simplices leads to the equations

\begin{eqnarray}\label{Gl:ReggeEquationsCurved}
\frac{\partial S_{\Tau}^{(\kappa)}}{\partial l_e}\;=\;\sum_{h\supset
e}\frac{\partial F^{(\kappa)}_h}{\partial
l_e}\epsilon_h^{(\kappa)}\;=\;0.
\end{eqnarray}

\subsection{Gauge invariance in Regge Calculus}

Analogously to our observations in the last chapter, the
reparametrization invariance of General Relativity is lost in Regge
Calculus, in the following sense: For a given set of boundary
lengths, the solutions to Regge's equations
(\ref{Gl:ReggeEquationsFlat}) are generically unique, i.e.
completely determined by the boundary data. The only exceptions to
this are the cases in which the discrete dynamics exactly reproduces
the continuum dynamics.

In $3D$ with $\Lambda=0$ the Regge equations
(\ref{Gl:ReggeEquationsFlat}) are simply the vanishing of the
deficit angles $\epsilon_e=0$, the solution of which is a
triangulation of a locally flat space-time. This is also the
solution to $3D$ GR with vanishing cosmological constant. In higher
dimensions there is, among other solutions, also $\epsilon_h=0$,
which can readily be seen to solve (\ref{Gl:ReggeEquationsFlat}) for
$\Lambda=0$. Again, this coincides with locally flat space-time
which is also one (among many solutions) of GR for $D>3$.

In all of these cases the solutions possess a vertex displacement
symmetry and an invariance under Pachner moves, which in $3D$ can
e.g. be seen as a result of the second Bianchi identities
\cite{HERBIEBIANCHI, FREILOU}. As a result, the bulk lengths
$l_e,\,e\in\mathcal{T}^\circ$ are not uniquely determined by the
boundary lengths $l_e,\,e\in\partial\mathcal{T}$, rather the vertex
displacement symmetry results in $D$ gauge degrees of freedom per
vertex.

Apart from these special cases, where the discrete dynamics exactly
reproduces the continuum dynamics, the boundary data fixes uniquely
the lengths of the edges in the interior of the triangulation
\cite{OURS}.\footnote{Apart from discrete ambiguities, which we
ignore for the time being \cite{PIRAN, OURS}.} That is translating a
vertex in a solution does not lead to another solution (as it does
for $3D$ Regge calculus with $\Lambda=0$) and the Hamilton--Jacobi
functional, i.e. the action evaluated on a solution, is not
invariant under Pachner moves of the bulk. This is analogous to the
situation in one dimension, where the reparametrization invariance
(which in discretized gravity would amount to an invariance under
change of triangulation) is lost in the na\"ive
discretization.\footnote{The exceptions for this, e.g. $3D$ with
$\Lambda=0$, can be compared to the free particle in one dimension,
where the continuum solutions are linear dependencies between the
$t_n$ and the $q_n$. In fact, the na\"ive discretization
(\ref{Gl:DiscretizationOfReparametrizationInvariantActionIn1D})
already coincides with the perfect action
(\ref{Gl:ExactDiscreteActionIn1D}) for this case, and the solutions
are not uniquely determined by the boundary data $t_0,t_N,q_0,q_N$.
Rather, the Hessian has as many zero Eigenvalues as inner vertices
and the solutions $t_n,q_n$ are non-unique in the sense of chapter
\ref{Ch:PerfectActionsIn1D}, which is a reflection of the gauge
symmetry of the continuum limit in this case.}

%
%
%
%

\section{Improved and perfect action in $3D$}

Since lattice gauge theory is not diffeomorphism-invariant, the
symmetries that are broken by discretization are not its local gauge
symmetries, which are of a different nature than in GR, but
Poincar\'e-invariance. The methods to construct improved and perfect
actions in QCD can therefore not be directly transferred to GR. We therefore attempt to generalize the way this is done for one-dimensional systems, encountered in chapter \ref{Ch:DiscretizedActionsIn1D}, to the case of Regge Calculus.

In one dimension the interval, on which the continuous theory is
defined, is divided into smaller intervals as a result of the
discretization, and in order to define the improved action
(\ref{Gl:ImprovedActionFor1D}) the interiors of these intervals are
then further refined. The discrete equations are then solved for the
refined lattice, subject to boundary conditions which relate them to
values on the coarse lattice. Therefore, since in Regge Calculus
spacetime is split into simplices via a triangulation, we will
refine this triangulation further into smaller simplices in order to
improve the action. Note that in more than one dimension the
boundary of a triangulation and between single simplices is
nontrivial, and it needs to be refined as well.

\subsection{Refinement of the Regge
action}\label{Ch:RefinementOfTheReggeAction}

We will first demonstrate the procedure for $D=3$ to show the
general idea, before we turn to the case of higher dimensions (in
particular $D=4$, which is the case of most interest
to us) in chapter \ref{Ch:ProcedureFor4D}.\\[5pt]

\begin{figure}[ht]
\begin{minipage}[b]{0.5\linewidth}
\centering
\includegraphics[scale=0.75]{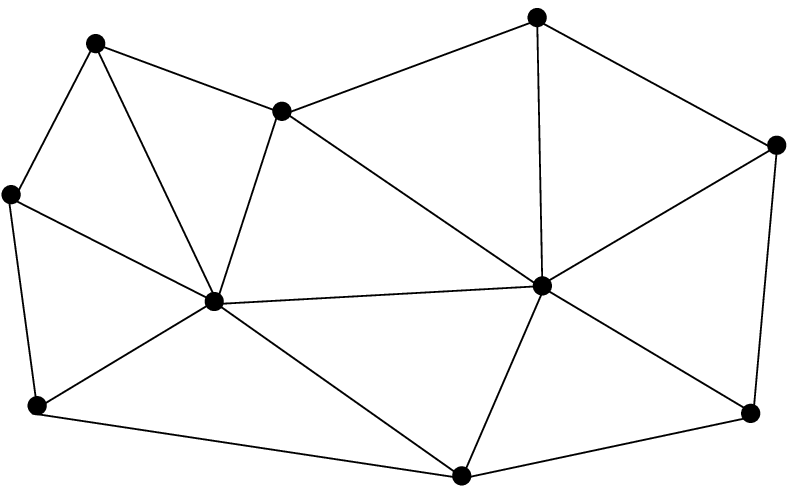}
\caption{\small Coarse triangulation $\Tau$ consisting of edged $E$,
triangles $T$ and tetrahedra $\Sigma$.}
\label{fig:CoarseTriangulation}
\end{minipage}
\hspace{0.5cm}
\begin{minipage}[b]{0.5\linewidth}
\centering
\includegraphics[scale=0.8]{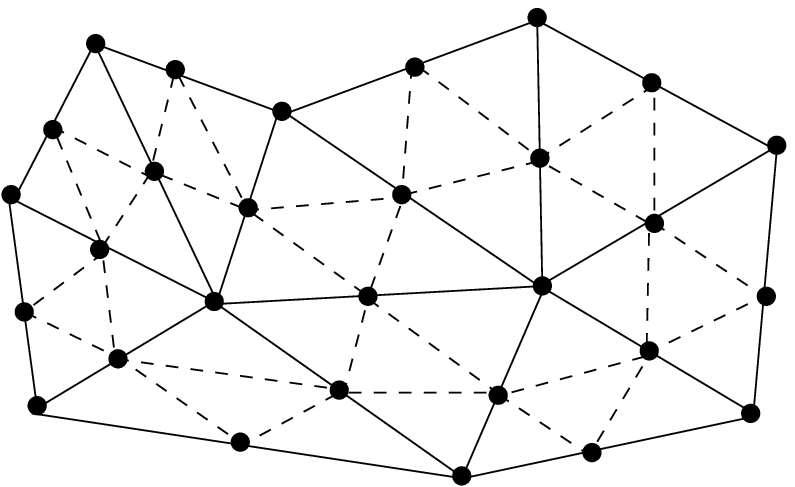}
\caption{Fine triangulation $\tau$ consisting of edged $e$,
triangles $t$ and tetrahedra $\sigma$.}
\label{fig:FineTriangulation}
\end{minipage}
\end{figure}

Consider a three-dimensional triangulation $\Tau$, consisting of
edges $E$, triangles $T$ and tetrahedra $\Sigma$, possibly with a
boundary $\partial\Tau$. The Regge action $S_\Tau$ is given by
(\ref{Gl:ReggeActionFlat}), and is a function of the edge lengths
$L_E$. Now subdivide $\Tau$ into a finer triangulation $\tau$,
consisting of edges $e$, triangles $t$ and tetrahedra $\sigma$.
Similarly to the definition of the improved action in $1D$, we solve
the Regge equations for the edge lengths $l_e$ subject to the
conditions
\begin{eqnarray}\label{Gl:ConditionsForLengthsIn3Dimensions}
\sum_{e\subset E}l_e\;=\;L_E
\end{eqnarray}

\noindent and define the improved action $S_{\Tau,\tau}$ as the
value of the Regge action $S_\tau$ on a solution of the equations
for $l_e$ subject to (\ref{Gl:ConditionsForLengthsIn3Dimensions}).
We add the constraint (\ref{Gl:ConditionsForLengthsIn3Dimensions})
via Lagrange multipliers, i.e. we have to vary the action
\begin{eqnarray}\label{Gl:3DReggeWithConstraint}
 S_\tau\;=\;\sum_el_e\,\varphi_e\;-\;\Lambda\sum_\sigma V_\sigma\;+\;\sum_E\alpha_E\left(L_E-\sum_{e\subset E}l_e\right)
\end{eqnarray}

\noindent where we have defined $\varphi_e:=\psi_e$ for
$e\in\partial\tau$ and $\varphi_e:=\epsilon_e$ for $e\in\tau^\circ$,
to unify notation. The equations of motion are then given by
deriving (\ref{Gl:3DReggeWithConstraint}) w.r.t the $l_e$ and
$\alpha_E$, i.e. one gets
\begin{eqnarray}\label{Gl:EquationForLIn3Dimensions}
\frac{\partial S_\tau}{\partial l_e}\;&=&\;\varphi_e\;-\;\Lambda\sum_{\sigma\supset e}\frac{\partial
V_\sigma}{\partial l_e}\;-\;\sum_{E\supset e}\alpha_E\;=\;0,\\[5pt]\label{Gl:EquationForAlphaIn3Dimensions}
\frac{\partial S_\tau}{\partial\alpha_E}\;&=&\;L_E-\sum_{e\subset E}l_e\;=\;0.
\end{eqnarray}

\noindent The improved action is then defined as the value of
$S_\tau$ on a solution of (\ref{Gl:EquationForLIn3Dimensions}),
(\ref{Gl:EquationForAlphaIn3Dimensions}), i.e.
\begin{eqnarray}
S_{\Tau,\tau}\;:=\;{S_\tau}_{\Big|{\frac{\partial S_\tau}{\partial l_e}=\frac{\partial S_\tau}{\partial\alpha_E}=0}}.
\end{eqnarray}

\noindent Note that the improved action $S_{\Tau,\tau}$ depends on
the ``large'' lengths $L_E$ for $E\in\Tau$, but incorporates the
dynamics of the finer triangulation $\tau$. A quick calculation
using Euler's theorem (\ref{Gl:EulersTheorem}) shows that the $l_e$,
$\alpha_E$ satisfy
\begin{eqnarray}
0\;=\;\sum_el_e\frac{\partial S_\tau}{\partial l_e}\;=\;\sum_el_e\varphi_e\;-\;3\Lambda\sum_\sigma V_\sigma\;-\;\sum_E\alpha_E\sum_{e\subset E}l_e.
\end{eqnarray}

\noindent So, with (\ref{Gl:EquationForAlphaIn3Dimensions}) the improved action can be put into the form
\begin{eqnarray}\label{Gl:ImprovedActionIn3D}
S_{\Tau,\tau}\;=\;\sum_EL_E\alpha_E\;+\;2\Lambda\sum_\Sigma V_\Sigma
\end{eqnarray}

\noindent where we have defined $
V_\Sigma:=\sum_{\sigma\subset\Sigma}V_\sigma$. Note that the
$\alpha_E,V_\Sigma$ are complicated functions of the $L_E$, which
have to be determined by the equations of motion
(\ref{Gl:EquationForLIn3Dimensions}),
(\ref{Gl:EquationForAlphaIn3Dimensions}). Nevertheless, one can
derive the equations of motion by varying $S_{\Tau,\tau}$ w.r.t. the
$L_E$. This can be achieved by changing the $L_E\to L_E+\delta L_E$,
and assuming that the solutions for the $l_e$ and $\alpha_E$ also
change only slightly to $l_e\to l_e+\delta l_e$,
$\alpha_E\to\alpha_E+\delta\alpha_E$. Therefore the value of
$S_{\Tau,\tau}$ changes by
\begin{eqnarray}
\delta S_{\Tau,\tau}\;&=&\;\sum_e\frac{\partial S_\tau}{\partial
l_e}\delta l_e\;+\;\sum_E\frac{\partial
S_\tau}{\partial\alpha_E}\delta\alpha_E\;+\;\sum_E\frac{\partial
S_\tau}{\partial L_E}\delta L_E\;_{\Big|{\frac{\partial
S_\tau}{\partial l_e}=\frac{\partial
S_\tau}{\partial\alpha_E}=0}}\\[5pt]\nonumber
&=&\;\alpha_E\,\delta L_E.
\end{eqnarray}

\noindent Therefore the equations for the $L_E$ determined by the
improved action $S_{\Tau,\tau}$ are the vanishing of the Lagrange
multipliers, i.e.
\begin{eqnarray}\label{Gl:EOMForImprovedActionIn3D}
\frac{\partial S_{\Tau,\tau}}{\partial L_E}\;=\;\alpha_E\;\stackrel{!}{=}\;0\qquad\text{for }E\in\Tau^\circ,
\end{eqnarray}

\noindent which, together with (\ref{Gl:EquationForLIn3Dimensions})
are equivalent to the Regge equations for the $l_e$\footnote{To be
precise: It is equivalent to $\frac{\partial S_\tau}{\partial
l_e}=0$ for all bulk edges $e\in\tau^\circ$, all boundary edges
$e\in\partial\tau$ which are \emph{not} a subedge of an edge in
$\partial\Tau$, i.e. $e\nsubseteq E$, and
(\ref{Gl:EquationForLIn3Dimensions}) for all $e\in\partial\tau$
which are, i.e. also $e\subset E\in\partial\Tau$. This is equivalent
to the Regge equations for the $e$ in the finer triangulation
$\tau$, plus the vanishing of the canonical momenta on the boundary
triangles $T\in\partial\Tau$.}.

\subsection{Perfect action in $3D$}

There is a similarity between the improved action (\ref{Gl:ImprovedActionIn3D}) and the Regge action with curved simplices (\ref{Gl:ReggeActionWithCurvedSimplices}), as well as the respective resulting equations of motion (\ref{Gl:EOMForImprovedActionIn3D}) and (\ref{Gl:ReggeEquationsCurved}). The similarity becomes more apparent if we define
\begin{eqnarray}\label{Gl:DefinitionOfLargeTheta}
\Theta_E^\Sigma:=\sum_{\sigma\supset
e,\,\sigma\in\Sigma}\left(\theta_e^\sigma-\Lambda\frac{\partial
V_\sigma}{\partial l_e}\right)
\end{eqnarray}

\noindent for some\footnote{The equations
(\ref{Gl:EquationForLIn3Dimensions}) guarantee that this choice does
not depend on the actual $e\subset E$.} $e\subset E$, then we have
$\alpha_E=2\pi-\sum_{\Sigma\supset E}\Theta_E^\Sigma$ for $E\subset
\Tau^\circ$ being in the bulk and $\alpha_E=\pi-\sum_{\Sigma\supset
E}\Theta_E^\Sigma$ for $E\subset\partial\Tau$ being in the boundary.
For every edge $E\subset\Tau^\circ$ in the bulk, the equations of
motion determined by $S_{\Tau,\tau}$ therefore are
\begin{eqnarray}\label{Gl:EOMForImprovedActionIn3D-2}
2\pi\,-\,\sum_{\Sigma\supset E}\Theta_E^\Sigma\;=\;0.
\end{eqnarray}

\noindent Note that, despite the formal similarity, the
$\Theta_E^\Sigma$ are not quite the interior dihedral angles at the
edges $E$ in the tetrahedra $\Sigma$. It is, however, not hard to
show that they become so in the perfect limit, i.e. the limit of
infinitely fine subdivision, which we denote by $\tau\to\infty$. If
the triangulation $\tau$ is such that its simplices $\sigma$ are
regular, i.e. their edge lengths $l_e$ after solving
(\ref{Gl:EquationForLIn3Dimensions}),
(\ref{Gl:EquationForAlphaIn3Dimensions}) are all of the same small
order of magnitude $\lambda$, then the term in
(\ref{Gl:DefinitionOfLargeTheta}) containing the derivative of the
volume scales like $O(\lambda^2)$, as compared to the
$\theta_e^{\sigma}$, which scale as $O(1)$, and therefore dominate
the expression. (Also note that $\theta^\sigma_e-\kappa
\frac{\partial V_\sigma}{\partial l_e}$ is the first order Taylor
expansion in $\kappa$ of the dihedral angle in a curved
tetrahedron.) We conclude that, in the perfect limit $\tau\to\infty$
the $\Theta_E^\Sigma$ indeed converge to the sum of the interior
angles at $e\subset E$ in $\sigma\subset\Sigma$, i.e. to the
interior angle at $E$ in $\Sigma$. Note that this interior angle is
the same everywhere ``on'' $E$, since it is independent of $e\subset
E$. Even more, we can demonstrate the perfect limit of $V_\Sigma$
(as a function of the $L_E$) by e.g. considering a triangulation
$\tau$ consisting of only one tetrahedron $|\Tau|=\Sigma$. Then the
variations (\ref{Gl:EOMForImprovedActionIn3D}) of $S_{\Tau,\tau}$
with respect to one of the $L_E$, $E=1,\ldots,6$ is equivalent to
\begin{eqnarray}
\sum_{E=1}^6L_E\frac{\partial \Theta_E}{\partial L_E}\;=\;2\Lambda
\frac{\partial V_\Sigma}{\partial L_E},
\end{eqnarray}

\noindent which, in the perfect limit, is exactly the Schlaefli
identity for curved tetrahedra (\ref{Gl:SchlaefliCurved}) which
related the interior angles of curved dihedral angles and volumes on
tetrahedra of constant curvature $\kappa=\Lambda$. In the perfect
limit, the formal similarity becomes an equality, and we conclude
that the perfect action in $3D$ is given by
\begin{eqnarray}\label{Gl:PerfectActionIn3D}
S_{\Tau,*}\;:=\;\lim_{\tau\to\infty}S_{\Tau,\tau}\;=\;S_{\Tau}^{(\kappa)},
\end{eqnarray}

\noindent i.e. coincides with the Regge action for constantly curved
tetrahedra with curvature $\kappa=\Lambda$. It is quite easy to show
that this action has three gauge degrees of freedom per vertex,
unlike $S_{\Tau}$, since the equations of motion - given by the
perfect limit of (\ref{Gl:EOMForImprovedActionIn3D-2}) - are
equivalent to the vanishing of all deficit angles $\epsilon_E=0$ for
interior edges $E\in\Tau^\circ$, which results in the triangulation
of a manifold of constant sectional curvature $\kappa=\Lambda$. This
not only reproduces exactly the continuum dynamics of $3D$ GR with
cosmological constant $\Lambda$, but also posesses the exact vertex
displacement symmetry as $3D$ Regge calculus with flat simplices
exhibits for $\Lambda=0$. Furthermore, it is invariant under
refinement of triangulation $\Tau$, as it should be by construction.

We conclude that in $3D$, the gauge symmetry of $GR$ containing $3$
gauge degrees of freedom per vertex, which is broken for
$\Lambda\neq 0$, is restored in the perfect limit. The Regge action
for constantly curved tetrahedra arises naturally as perfect action
in this context. It should be noted that the Regge action
(\ref{Gl:ReggeActionFlat}) with flat simplices arises naturally as
first order approximation, by the following argument: By
investigating the scaling property of the curved Regge action
(\ref{Gl:ReggeActionWithCurvedSimplices}), e.g. by considering
(\ref{Gl:ScalingOfVolume}), one can easily see that a scaling of the
edge lengths $l_e\to\lambda l_e$ can be absorbed into a scaling of
the curvature $\kappa\to\lambda^2\kappa$. Expanding the curved
functions $\theta_e^{(\kappa), \sigma}$, $V_\sigma^{(\kappa)}$ into
linear order in $\kappa$, one obtains, by using the identities
(\ref{Gl:DerivationOfAngle}) and (\ref{Gl:SchlaefliCurved}), that
\begin{eqnarray}
S_{\tau}^{(\kappa)}\;=\;S_\tau\;+\;O(\kappa^2)
\end{eqnarray}

\noindent where $S_\tau$ is the Regge action (\ref{Gl:ReggeActionFlat}) for flat simplices with cosmological constant $\Lambda=\kappa$.

\section{Higher dimensions}\label{Ch:ProcedureFor4D}

We now consider the concept of improved and perfect actions for
dimensions $D>3$, where of course the case of ultimate interest is
$D=4$. Nevertheless, since the arising procedures are generic for
arbitrary higher dimensions, we shall treat the problem for
arbitrary dimension $D$, and comment about the implications for
$D=4$ in the end.

The general concept for defining the improved action $S_{\Tau,\tau}$
for $D>3$ is similar to $D=3$. We start with a triangulation $\Tau$
consisting of simplices $\Sigma$, hinges $H$ and edges $E$. Now
subdivide $\Tau$ into a finer triangulation $\tau$, consisting of
$D$-simplices $\sigma$, hinges $h$ and edges $e$. Note that some of
the hinges $h$ are contained in the ``larger'' hinges $H$. The
action for the finer triangulation $S_{\tau}$ is a function of the
edge lengths $l_e$. It turns out that the most convenient
generalization of the condition
(\ref{Gl:ConditionsForLengthsIn3Dimensions}) to $D>3$ is not to keep
the edge lengths $L_E$ fixed, but rather the $D-2$-volumes $F_H$,
i.e. to constrain the variation of the Regge action for $\tau$ by
\begin{eqnarray}\label{Gl:ConditionsForLengthsInDDimensions}
\sum_{h\subset H}f_h\;=\;F_H,
\end{eqnarray}

\noindent where $f_h$ is the $D-2$-volume of the hinge $h$. In other words, we vary
\begin{eqnarray}\label{Gl:ReggeWithConstraint}
 S_\tau\;=\;\sum_hf_h\,\varphi_h\;-\;\Lambda\sum_\sigma V_\sigma\;+\;\sum_H\alpha_H\left(F_H-\sum_{h\subset H}f_h\right)
\end{eqnarray}

\noindent with respect to $l_e$ and $\alpha_H$, where the Lagrange
multipliers $\alpha_H$ have been introduced in order to enforce
(\ref{Gl:ConditionsForLengthsInDDimensions}), and $\varphi_h$
denotes the deficit angle $\epsilon_h$ for $h\in\tau^\circ$ being in
the bulk, and the extrinsic curvature angle $\psi_h$ for
$h\in\partial \tau$ in the boundary. The improved action is -
similar as for $D=3$ - defined as
\begin{eqnarray}
S_{\Tau,\tau}\;:=\;{S_\tau}_{\Big|{\frac{\partial S_\tau}{\partial l_e}=\frac{\partial S_\tau}{\partial\alpha_H}=0}},
\end{eqnarray}

\noindent and is naturally a function of the $F_H$ (e.g. the areas
of the triangles for $D=4$.). The resulting equations for the
$l_e,\alpha_H$ are, using the Schlaefli-identity
(\ref{Gl:SchlaefliFlat})
\begin{eqnarray}\label{Gl:EquationForLInDDimensions}
\frac{\partial S_\tau}{\partial l_e}\;&=&\;\sum_{h\supset
e}\frac{\partial f_h}{\partial
l_e}\varphi_h\;-\;\Lambda\sum_{\sigma\supset e}\frac{\partial
V_\sigma}{\partial l_e}\;-\;\sum_{h\supset e}\sum_{H\supset
h}\alpha_H\frac{\partial f_h}{\partial
l_e}\;=\;0,\\[5pt]\label{Gl:EquationForAlphaInDDimensions}
\frac{\partial S_\tau}{\partial\alpha_H}\;&=&\;F_H-\sum_{h\subset H}f_h\;=\;0.
\end{eqnarray}

\noindent Using Euler's theorem (\ref{Gl:EulersTheorem}) we get
\begin{eqnarray}
 0=\sum_{e}l_e\frac{\partial S_\tau}{\partial l_e}\;=\;(D-2)\sum_hf_h\varphi_h\;-\;D\Lambda\sum_\sigma V_\sigma\;-\;(D-2)\sum_H\alpha_H\sum_{h\subset H}f_h,
\end{eqnarray}

\noindent which, inserted into (\ref{Gl:ReggeWithConstraint})
together with (\ref{Gl:EquationForAlphaInDDimensions}), results in
the improved action
\begin{eqnarray}\label{Gl:ImprovedActionInDDimensions}
 S_*\;=\;\sum_HF_H\alpha_H\;+\;\frac{2}{D-2}\Lambda\sum_\Sigma V_\Sigma,
\end{eqnarray}

\noindent where we have defined
$V_\Sigma:=\sum_{\sigma\subset\Sigma}V_\sigma$. Note the similarity
between the improved action (\ref{Gl:ImprovedActionInDDimensions})
and the Regge action (\ref{Gl:ReggeActionWithCurvedSimplices}) for
simplices of constant curvature $\kappa$, for $\kappa$ and $\Lambda$
being related by (\ref{Gl:LambdaAndKappaRelation}).

The improved action (\ref{Gl:ImprovedActionInDDimensions}) is a
function of the $F_H$ via the $\alpha_H$ and $V_\Sigma$, which will
depend on the $F_H$ in a complicated manner to be determined by
solving the equations (\ref{Gl:EquationForLInDDimensions}),
(\ref{Gl:EquationForAlphaInDDimensions}). Nevertheless, we can
derive the equations for the $F_H$ determined by the improved
action. For this we consider the same set of equations, just with
slightly changed parameters $F_H+\delta F_H$. It can be expected
that the solutions for $l_e,\alpha_H$ will also change just slightly
via
\begin{eqnarray}\nonumber
l_e\;&\longrightarrow&\;l_e\,+\,\delta l_e\\[5pt]\nonumber
\alpha_H\;&\longrightarrow&\;\alpha_{H}\,+\,\delta \alpha_{H}
\end{eqnarray}

\noindent Then $S_{\Tau,\tau}$ changes slightly via
\begin{eqnarray}\label{Gl:VariationOfAction}
\delta S_{\Tau,\tau}\;&=&\;\sum_e\frac{\partial S_\tau}{\partial l_e}\delta
l_e\;+\;\sum_H\frac{\partial S_\tau}{\partial
\alpha_H}\delta\alpha_H\;+\;\sum_H\frac{\partial S_\tau}{\partial
F_H}\delta F_H,
\end{eqnarray}

\noindent and evaluating (\ref{Gl:VariationOfAction}) on a solution results in
\begin{eqnarray}
\frac{\partial S_{\Tau,\tau}}{\partial F_H}\;=\;\alpha_H\;=\;0.
\end{eqnarray}

\subsection{Improving the curved Regge action}

It is instructive to repeat the calculation with curved simplices.
We start from the action (\ref{Gl:ReggeActionWithCurvedSimplices})
and impose the constraints via Lagrange multipliers
$\alpha_H^{(\kappa)}$. In other words, we have to vary the action
\begin{eqnarray}\label{Gl:ReggeActionWithConstraintsCurved}
S_\tau^{(\kappa)}\;=\;\sum_h f_h^{(\kappa)}\varphi_h^{(\kappa)}\;+\;(D-1)\kappa\sum_\sigma V_\sigma^{(\kappa)}\;+\;\sum_H\alpha_H^{(\kappa)}\left(F_H-\sum_{h\subset H}f_h^{(\kappa)}\right)
\end{eqnarray}

\noindent where the superscript ${}^{(\kappa)}$ denotes the volume
of hinges and simplices of constant curvature $\kappa$. Again,
$\varphi_h^{(\kappa)}$ is shorthand for $\epsilon_h^{(\kappa)}$
whenever $h\in\tau^\circ$ is a hinge in the bulk, and
$\psi_h^{(\kappa)}$, whenever $h\in\partial\tau$ is in the boundary.
With the Schlaefli-identity (\ref{Gl:SchlaefliCurved}) for simplices
of constant curvature, the resulting equations for the $l_e$ are
\begin{eqnarray}\label{Gl:EquationForLInDDimensionsCurved}
\frac{\partial S_\tau^{(\kappa)}}{\partial l_e}\;&=&\;\sum_{h\supset
e}\frac{\partial f_h^{(\kappa)}}{\partial
l_e}\varphi_h^{(\kappa)}\;-\;\sum_{h\supset e}\sum_{H\supset
h}\alpha_H^{(\kappa)}\frac{\partial f_h^{(\kappa)}}{\partial
l_e}\;=\;0,\\[5pt]\label{Gl:EquationForAlphaInDDimensionsCurved}
\frac{\partial S_\tau^{(\kappa)}}{\partial\alpha_H^{(\kappa)}}\;&=&\;F_H-\sum_{h\subset H}f_h^{(\kappa)}\;=\;0.
\end{eqnarray}

\noindent With the geometric identity (\ref{Gl:MainFormula}) for simplices of constant curvature $\kappa$, we get
\begin{eqnarray}
0\;=\;\sum_{e}l_e\frac{\partial S_\tau^{(\kappa)}}{\partial
l_e}\;&=&\;(D-2)\sum_{h}f_h^{(\kappa)}\varphi_h^{(\kappa)}\;+\;2\kappa\sum_h\frac{\partial
f_h^{(\kappa)}}{\partial \kappa}\varphi_h^{(\kappa)}\\[5pt]\nonumber
&&\;-(D-2)\sum_H\sum_{h\subset H}\alpha_H^{(\kappa)} f_h^{(\kappa)}\;-\;2\kappa\sum_H\sum_{h\subset H}\alpha_H^{(\kappa)} \frac{\partial f_h^{(\kappa)}}{\partial \kappa}
\end{eqnarray}

\noindent which results in the improved action
\begin{eqnarray}\nonumber
S_{\Tau,\tau}^{(\kappa)}\;&=&\;\sum_HF_H\alpha_H^{(\kappa)}\;+\;(D-1)\kappa\sum_\Sigma V_\Sigma^{(\kappa)}\\[5pt]\label{Gl:ImprovedActionCurved}
&&\;+\;\frac{2}{D-2}\kappa\sum_h\frac{\partial
f_h^{(\kappa)}}{\partial
\kappa}\varphi_h^{(\kappa)}\;+\;\frac{2}{D-2}\kappa\sum_H\sum_{h\subset
H}\alpha_H^{(\kappa)} \frac{\partial f_h^{(\kappa)}}{\partial
\kappa}
\end{eqnarray}

\noindent where we have defined
$V_\Sigma^{(\kappa)}:=\sum_{\sigma\subset\Sigma}V_\sigma^{(\kappa)}$.
By a similar reasoning as in the case with flat simplices, the
equations for the improved action is easily obtained to be
\begin{eqnarray}
\frac{\partial S_{\Tau,\tau}^{(\kappa)}}{\partial F_H}\;=\;\alpha_H^{(\kappa)}\;=\;0.
\end{eqnarray}

\subsection{Perfect actions with flat and curved simplices}

If we consider the improved actions
(\ref{Gl:ImprovedActionInDDimensions}) and
(\ref{Gl:ImprovedActionCurved}), which result from refining the
triangulations with flat and curved simplices, respectively, we see
that their expressions seem to be quite different. However, in
performing the continuum limit for both actions, we will demonstrate
that they both converge to the same perfect action, when $\Lambda$
and $\kappa$ satisfy the relation (\ref{Gl:LambdaAndKappaRelation}).
In order to do this, we show that - as functions of the lengths
$F_H$ - both perfect limits satisfy the same ordinary differential
equation w.r.t. $\Lambda$ (or, equivalently, $\kappa$). We do this
by considering the ODE's that the two improved actions
(\ref{Gl:ImprovedActionInDDimensions}) and
(\ref{Gl:ImprovedActionCurved}) satisfy, and show that in the
continuum limit they converge to each other.

We first vary the improved action $S_{\Tau,\tau}$ for flat simplices w.r.t
$\Lambda$, by solving the equations of motion again with
$\Lambda\;\rightarrow\;\Lambda+\delta\Lambda$, and assume the
resulting solutions $l_e,\alpha_H$ also change only slightly by
$l_e\to l_e+\delta l_e$ and $\alpha_H\to\alpha_H+\delta \alpha_H$.
The change of the action is therefore
\begin{eqnarray}\nonumber
\delta S_{\Tau,\tau}\;&=&\;\sum_e\frac{\partial S_\tau}{\partial l_e}\delta
l_e\;+\;\sum_H\frac{\partial S_\tau}{\partial
\alpha_H}\delta\alpha_H\;+\;\frac{\partial S_\tau}{\partial
\Lambda}\delta \Lambda\\[5pt]
&=&\; -\sum_\Sigma V_\Sigma\;\delta \Lambda,
\end{eqnarray}

\noindent where the Regge equations have been used. With
(\ref{Gl:ImprovedActionInDDimensions}) this results in
\begin{eqnarray}\label{Gl:EOMForImprovedActionFlatSimplices}
S_{\Tau,\tau}\;+\;\frac{2}{D-2}\Lambda\frac{\partial S_{\Tau,\tau}}{\partial
\Lambda}\;=\;\sum_{H} F_H\alpha_H.
\end{eqnarray}

\noindent The same calculation for the improved action
(\ref{Gl:ImprovedActionCurved}) with curved simplices is more
involved, since the constituents depend explicitly on $\kappa$.
Since $S_{\Tau,\tau}^{(\kappa)}$ is the value of $S_\tau^{(\kappa)}$
evaluated on a solution, varying $S_{\Tau,\tau}^{(\kappa)}$ w.r.t
$\kappa$ is equivalent to varying $S_\tau^{(\kappa)}$, and inserting
the solutions for $l_e,\alpha_H^{(\kappa)}$ afterwards (since the
variations of $l_e,\alpha_H^{(\kappa)}$ vanish on solutions, by
definition). We have
\begin{eqnarray}
\frac{\partial S_{\Tau,\tau}^{(\kappa)}}{\partial \kappa}\;&=&\;\frac{\partial
S_\tau^{(\kappa)}}{\partial \kappa}\;=\;\sum_h\frac{\partial
f_h^{(\kappa)}}{\partial\kappa}\varphi_h^{(\kappa)}\;+\;\sum_hf_h^{(\kappa)}\frac{\partial
\epsilon_h^{(\kappa)}}{\partial\kappa}\;+\;(D-1)\sum_\sigma
V_\sigma^{(\kappa)}\\[5pt]\nonumber
&&\;+\;(D-1)\kappa\sum_\sigma\frac{\partial V_\sigma^{(\kappa)}}{\partial\kappa}\;-\;\sum_H\sum_{h\subset H}\alpha_H^{(\kappa)}\frac{\partial f_h^{(\kappa)}}{\partial\kappa}
\end{eqnarray}

\noindent With (\ref{Gl:DerivationOfAngle}) and the Schlaefli
identity (\ref{Gl:SchlaefliCurved}), we have
\begin{eqnarray}
\sum_hf_h^{(\kappa)}\frac{\varphi_h^{(\kappa)}}{\partial\kappa}\;=\;-\frac{D-1}{2}\sum_\sigma\sum_{e\subset\sigma}l_e\frac{\partial
V_\sigma^{(\kappa)}}{\partial l_e}
\end{eqnarray}

\noindent which results in
\begin{eqnarray}
S_\tau^{(\kappa)}\;+\;\frac{2}{D-2}\kappa\frac{\partial
S_\tau^{(\kappa)}}{\partial \kappa}\;=\;\sum_h\sum_{e\subset
h}l_e\frac{\partial f_h^{(\kappa)}}{\partial
l_e}\varphi_h^{(\kappa)}\;+\;\sum_{H}\alpha_H^{(\kappa)}F_H\;-\;\sum_{h\subset
H}\sum_{e\subset h}l_e\frac{\partial f_h^{(\kappa)}}{\partial l_e}
\end{eqnarray}

\noindent which, evaluated on a solution to
(\ref{Gl:EquationForLInDDimensionsCurved}),
(\ref{Gl:EquationForAlphaInDDimensionsCurved}) results in
\begin{eqnarray}\label{Gl:EOMForImprovedActionCurvedSimplices}
S_{\Tau,\tau}^{(\kappa)}\;+\;\frac{2}{D-2}\kappa\frac{\partial
S_{\Tau,\tau}^{(\kappa)}}{\partial \kappa}\;=\;\sum_{H}\alpha_H^{(\kappa)}F_H.
\end{eqnarray}

\noindent Note the similarity to
(\ref{Gl:EOMForImprovedActionFlatSimplices}).\\[5pt]

The solutions for the $\alpha_H,\,\alpha_H^{(\kappa)}$ in fact
converge to each other in the perfect limit. In order to show this,
we assume that $l_e,\alpha_H$ satisfy the equations
(\ref{Gl:EquationForLInDDimensions}),
(\ref{Gl:EquationForAlphaInDDimensions}), and $l_e+\delta
l_e,\alpha_H+\delta \alpha_H$ satisfy the equations
(\ref{Gl:EquationForLInDDimensionsCurved}),
(\ref{Gl:EquationForAlphaInDDimensionsCurved}). We consider the
limit of very fine triangulations $\tau$ - in particular we assume
that both solutions are sufficiently close to a solution to the
Einstein equations - this in particular means that the scale over
which the curvature changes is much larger than $l_e$ or $l_e+\delta
l_e$. For curved simplices the limit of small edge lengths coincides
with the limit of small curvature. Expanding curved quantities in
$\kappa$ results in
\begin{eqnarray}
V_\sigma^{(\kappa)}\;=\;V_\sigma\;+\;\kappa\frac{\partial
V^{(\kappa)}_\sigma}{\partial\kappa}_{|_{\kappa=0}}\;+\;O(\kappa^{2})
\end{eqnarray}

\noindent In the appendix \ref{Ch:AppendixGeometricIdentity} it is
proved that the term linear in $\kappa$ is of order $O(l_e^{D+2})$.
Furthermore, for a dihedral angle $\theta_h^{\sigma\,(\kappa)}$ one
has, using (\ref{Gl:DerivationOfAngle}) and the Schlaefli identity
(\ref{Gl:SchlaefliCurved})
\begin{eqnarray}
\sum_{h\subset\sigma}f_h^{(\kappa)}\theta_h^{\sigma\,(\kappa)}
\;=\;\sum_{h\subset\sigma}f_h\theta_h^{\sigma}\;+\;\frac{D(D-1)}{2}\kappa\,V_\sigma\;+\;\kappa\sum_{h\subset\sigma}\frac{\partial
f_h^{(\kappa)}}{\partial
\kappa}_{|_{\kappa=0}}\theta_h^{\sigma}\;+\,O(\kappa^{2})
\end{eqnarray}

\noindent where quantities without superscript are volumes and angles
in flat simplices. As a result, we get
\begin{eqnarray}
S_\tau^{(\kappa)}\;=\;S_\tau\;+\;\kappa\sum_h\frac{\partial
f_h^{(\kappa)}}{\partial \kappa}_{|_{\kappa=0}}\xi_h\,+\,O(\kappa^2)
\end{eqnarray}

\noindent where
\begin{eqnarray}
\xi_h\;:=\;\left\{ \begin{array}{lcr}\displaystyle\varphi_h-\sum_{H\supset
h}\alpha_H&\text{for}&h\subset H\\\varphi_h&\text{for}& h\nsubseteq
H\end{array}\right.
\end{eqnarray}

\noindent Due to the Regge equations
(\ref{Gl:EquationForLInDDimensions})
\begin{eqnarray}
\sum_{h\supset e}\frac{\partial f_h}{\partial
l_e}\,\xi_h\;=\;\Lambda\sum_{\sigma\supset e}\frac{\partial
V_\sigma}{\partial l_e}
\end{eqnarray}

\noindent and due to the assumed regularity of the triangulation
$\tau$, where the edge lengths are all of the order of magnitude of
some lengths $l$, one has that $\xi_h\sim l^2$. In the limit of very
fine $\tau$, both $l_e$ and $l_e+\delta l_e$ can expected to tend to
zero, so we can expand (\ref{Gl:EquationForLInDDimensionsCurved}) in
$\delta l_e$ and compare it with $l_e$. We get
\begin{eqnarray}
\sum_{e'}\frac{\partial^2 S_\tau}{\partial l_e\partial
l_{e'}}\;\delta l_{e'}\;+\;\sum_H\frac{\partial^2S_\tau}{\partial
l_e\partial
\alpha_H}\delta\alpha_H\;+\;\kappa\frac{\partial}{\partial
l_e}\sum_h\frac{\partial
f^{(\kappa)}_h}{\partial\kappa}_{|_\kappa=0}\xi_h\;+\;O(l^{D+3})
\end{eqnarray}

\noindent Since $\frac{\partial^2 S_\tau}{\partial l_e\partial
l_{e'}}\sim l^{D-2}$ and $\frac{\partial^2S_\tau}{\partial
l_e\partial \alpha_H}\sim l$ for $e\subset H$, we get that
\begin{eqnarray*}
\delta l_{e'}\;&\sim&\;l^3\\[5pt]
\delta\alpha_H\;&\sim&\;l^D.
\end{eqnarray*}

\noindent Hence the perfect limit $\tau\to\infty$ corresponds to the
limit $l\to 0$. Therefore
$\alpha_H^{(\kappa)}=\alpha_H+\delta\alpha_H$ converges to
$\alpha_H$ in the continuum limit.\\[5pt]

Furthermore, the perfect actions $S_{\Tau,*}$ and
$S_{\Tau,*}^{(\kappa)}$ obviously coincide for $\kappa=\Lambda=0$.
So not only do they satisfy the same ODE w.r.t.
$\Lambda=(D-1)(D-2)\kappa/2$, which is first order, they also
coincide for one value. Therefore, they must coincide as functions
of the $F_H$, and we conclude
\begin{eqnarray}\label{Gl:FlatAndCurvedPerfectActionsCoincide}
S_{\Tau,*}\;=\;\lim_{\tau\to\infty}S_{\Tau,\tau}\;=\;\lim_{\tau\to\infty}S^{(\kappa)}_{\Tau,\tau}\;=\;S_{\Tau,*}^{(\kappa)}.
\end{eqnarray}

\subsection{Constantly curved subsector}

For $D>3$ it is nontrivial to compute the perfect limit of the
improved action $S_{\Tau,\tau}$ given by
(\ref{Gl:ImprovedActionInDDimensions}), since the $\alpha_H$ do not
necessarily, unlike in $D=3$, have to have the interpretation of
deficit angles at the hinges $H$ in that limit. In general, it will
be quite complicated to compute the $\alpha_H$ in general. However,
there is a special case in which one can compute the perfect action
$S_{\Tau,*}$, which is when the $F_H$ satisfy the following
requirement:

Let $\Tau$ be a triangulation of a manifold $|\Tau|=M$ with constant
curvature $\kappa$ with constantly curved simplices $\Sigma$, such
that there are vanishing deficit angles. If the $D-2$-dimensional
hinges $H$ have a volume $F_H$, then the value of the perfect action
$S_{\tau,*}$ on that configuration $F_H$ is given by
\begin{eqnarray}\label{Gl:PerfectActionInDDimensionsConstantlyCurved}
S_{\Tau,*}(F_H)\;=\;\sum_{H\subset\partial\Tau}F_H\left(\pi-\sum_{\Sigma\supset H}\theta_{H}^{(\kappa)\,\Sigma}\right)\;+\;(D-1)\kappa V_{M},
\end{eqnarray}

\noindent where $\theta_{H}^{(\kappa)\,\Sigma}$ is the dihedral
angle in the curved simplex $\Sigma$ at the hinge $H$, and $V_M$ is
the volume of the manifold $M$. This can be seen as follows: In the
last section we have shown that the Regge action with curved
simplices and the flat simplices lead to the same perfect action
$S_{\Tau,*}$ if $\Lambda$ and $\kappa$ are related by
(\ref{Gl:LambdaAndKappaRelation}). Therefore we can use curved
simplices instead of flat ones in our triangulation $\Tau$. However,
curved simplices can be glued together with vanishing deficit angles
$\epsilon_H^{(\kappa)}=0$ to form the manifold $M$, since $M$ has
constant sectional curvature $\kappa$. There are in fact infinitely
many ways to do this, which can all be related by Pachner moves that
do not change the boundary $\partial\Tau$. For all of these
possibilities, the geometry satisfies trivially the Regge equations
(\ref{Gl:ReggeEquationsCurved}), because the deficit angles all
vanish. Moreover, the constraints
(\ref{Gl:ConditionsForLengthsInDDimensions}) are satisfied by
definition. The value of the Regge action $S_{\Tau}$ does not
actually depend on the exact triangulation $\Tau$, it is only
depending on the boundary data, i.e. the $F_H$ for
$H\in\partial\Tau$ and the extrinsic dihedral angles. The action
(\ref{Gl:ReggeActionWithCurvedSimplices}) evaluated on
$\epsilon_H^{(\kappa)}=0$ gives exactly
(\ref{Gl:PerfectActionInDDimensionsConstantlyCurved}). Since it is
invariant under refinement of the triangulation, it is by definition
the perfect action. Moreover, it is invariant under Pachner moves,
and invariant under variations of the $F_H$ which result of vertex
displacements, since these only change the $F_H$ in the
triangulation, but do not change the geometry, which is that of
constant curvature. Thus, in this special case we regain $D$ gauge
degrees of freedom per vertex (the vertex displacements), which
reflects the diffeomorphism symmetry of lapse and shift from the
continuum theory.

Note that in this case (\ref{Gl:ImprovedActionCurved}) shows that
$\alpha_H\equiv \epsilon_H^{(\kappa)}$ for $H\in\Tau^\circ$.
Moreover, for the special case of $\Tau$ consisting of one simplex
$\Sigma$, we can - in a similar derivation as for $D=3$, show that
\begin{eqnarray}
 \sum_HF_H\frac{\partial \alpha_H}{\partial F_{H'}}\;=\;(D-1)\kappa\frac{\partial V_{\Sigma}}{\partial F_{H'}},
\end{eqnarray}

\noindent which - since the $\frac{D(D-1)}{2}\times
\frac{D(D-1)}{2}$-matrix $\partial F_H/\partial L_E$ is
invertible\footnote{Apart from discretely many cases, see e.g.
\cite{barrett}.} - is equivalent to the Schlaefli identity within
curved simplices (\ref{Gl:SchlaefliCurved}).

In general, the $F_H$ that are the arguments of the improved and the
perfect action will not satisfy the requirement that there exists a
triangulation of curved simplices that can be glued together with
vanishing deficit angles\footnote{It might not be possible to glue
constantly curved simplices with these $F_H$ together at all -
although for each separate simplex the relation between the $L_E$
and the $F_H$ can be inverted, the resulting geometries of
neighboring simplices might be incompatible. One can suspect that
the geometry described will not be that of constantly curved
simplices, but rather of objects which are topological simplices,
but have a geometry which satisfies Einstein's equations in $D$
dimensions with a cosmological constant $\Lambda$ (of which the
constantly curved ones are a special case).}. In these cases
$\alpha_H$ will have a much more complicated interpretation, and
will be much harder to compute. In the case above where we have
computed the perfect action, however, we have recovered the perfect
action to reproduce a manifold with constant curvature $\kappa$,
which is a solution of the continuum theory of GR, which exists in
all dimensions $D$, as we have shown in chapter
(\ref{Ch:ContinuousPreliminaries}). For $D>3$, the sector of
solutions is much larger, however, and contains many more solutions.

\section{Summary and Conclusion}

We have investigated the concept of improved and perfect actions in
Regge calculus, where the reparametrization invariance of General
Relativity is usually broken.

Discretizations of theories with symmetries usually lose that
symmetry, e.g. in lattice gauge theory, where Poincar\'e-invariance
is broken by introduction of a lattice. The motivation for our
analysis was that the concept of improved and perfect actions is
used in order to regain the symmetry within the lattice formulation.
The QCD Lagrangian is not diffeomorphism invariant, however, and the
techniques for lattice QCD are therefore not directly applicable to
Regge Gravity.

It is well-known that one-dimensional systems with reparametrization
invariance lose that symmetry upon discretization, and there is a
procedure to construct improved and perfect actions in this case in
order to arrive at discrete actions which retain exact
reparametrization invariance \cite{marsden}. We have reviewed this
in chapter \ref{Ch:DiscretizedActionsIn1D}, and have proposed a
procedure to construct improved and perfect (classical) actions for
discretized, reparametrization invariant field theories, in
particular Regge Calculus in chapter
\ref{Ch:RefinementOfTheReggeAction}. We have applied this scheme to
Regge gravity in arbitrary dimensions $D$.\\[5pt]

We have done this by considering improved actions $S_{\Tau,\tau}$
which are defined on a triangulation $\Tau$, which however
incorporate the dynamics of the refined triangulation $\tau$, i.e.
is closer to the actual continuum dynamics.  In the canonical
formulation this leads also to a better approximation of  the
constraints in the sense of \cite{OURS}. It seems that this is a
useful setting in which to think about renormalization group flow in
a diffeomorphism-invariant context. Since the actual scales in the
theory have to be determined dynamically, they are not available to
label a cut-off for the theory, and in particular to compare them
for different labels. However, one can investigate the difference of
the dynamics which are discretized on two triangulations $\Tau$ and
$\tau$, the continuum limit being better and better approximated the
larger the difference between the two, i.e. the finer $\tau$ is
compared to $\Tau$.\footnote{Since the triangulations $\tau$ form a
partially ordered set, it might be that - in mathematical terms -
the renormalization group flow in this context has to be treated
with the convergence of filters, rather than sequences.}\\[5pt]

We have shown that the perfect action or $3D$ Regge
calculus for $\Lambda\neq 0$ can be computed
explicitly.\footnote{For $\Lambda=0$ the Regge action is already
perfect.}  It can be obtained by replacing the flat tetrahedra by
tetrahedra of constant curvature $\kappa=\Lambda$. This leads to the
action (\ref{Gl:PerfectActionIn3D}) which exactly reproduces the
continuum dynamics of $3D$ GR with cosmological constant, i.e.
vanishing deficit angles $\epsilon^{(\kappa)}_e=0$, leading to
space-time with constant local curvature. As a consequence, the thus
obtained perfect action leads to a similar vertex displacement
symmetry than one finds in $3D$ Regge calculus for $\Lambda=0$.

Since for $D>3$ the continuum theory possesses local degrees of
freedom, the perfect action is much harder to construct in this
case. Nevertheless, we could show that the Regge actions with flat
simplices, and that with simplices of constant curvature
$\kappa=\Lambda/3$ lie in the same universality class, i.e. lead to
the same perfect action $S_{\Tau,*}$. Moreover, we were able to
express the perfect action  in terms of the continuum limit of the
Lagrange multipliers $\alpha_H$ and the volumes of the simplices.
For the subsector of constantly curved solutions, which exists in GR
for all dimensions, the $\alpha_H$ can in fact be computed to be the
deficit- (or, in case of boundary hinges $H\subset\partial\Tau$,
extrinsic-) angles in constantly curved simplices, where the
curvature and the cosmological constant are related by
(\ref{Gl:LambdaAndKappaRelation}). For this subsector of solutions,
the perfect action possesses the vertex displacement symmetry, which
lead to $D$ gauge degrees of freedom per vertex. It therefore
captures the gauge symmetry of lapse and shift, since it reproduces
exactly the continuum dynamics (of constant curvature).

In this work we did not obtain explicitly an improved action which
takes into account propagating degrees of freedom. This would
correspond to integrating out higher frequency gravitons and their
interactions and finding an effective action. We expect this to be a
very complicated task leading to a non--local action. However it is
a promising one with possible contacts to other quantum gravity
approaches \cite{asymptsafety}. As a first step one can consider an
expansion around flat space and define an action that takes into
account the lowest non--linear dynamics of the gravitons
\cite{toappear}. As the perfect action is by construction
triangulation independent this could be also helpful for
understanding how to obtain triangulation independent models.
\\[5pt]

The $\kappa$--curved simplices, which appear in the improved and perfect actions, can be useful for the construction of quantum gravity models for several reasons.

\begin{itemize}
\item Using the perfect action $S_\Tau^{(\kappa)}$ given by
(\ref{Gl:PerfectActionIn3D}) instead of the Regge action
(\ref{Gl:ReggeActionFlat}), is a more appropriate description for
the problem at hand, since for $3D$ the perfect action correctly
reflects the finite number of degrees of freedom of the continuum
theory. These are not directly visible if one uses flat tetrahedra,
since for $\Lambda\neq 0$ the corresponding Regge equations lead to
a unique solution for the edge lengths. So no gauge freedom is
apparent in this description. The edge lengths can therefore be
mistaken to be physical degrees of freedom. The perfect action
however is not only invariant under further refinement of the
triangulation, it also shows that the edge lengths in itself are not
physical, but rather are a gauge artefact introduced by a choice of
triangulation.

Not only shows this that in construction of quantized models of $3D$
Regge calculus with $\Lambda\neq 0$ the perfect action
$S_\Tau^{(\kappa)}$ might be more suitable than $S_\Tau$, in a
broader context it shows how in discretized gravity theories it can
be difficult to tell physical from gauge degrees of freedom.

This is in particular important in $4D$, where the solutions to the
Regge equations (even for $\Lambda=0$) are generically unique. This
is usually taken as proof that the diffeomorphism symmetry of GR has
been successfully divided out, and one is only working with
gauge-invariant quantities (i.e. the edge lengths), since the gauge
symmetry of GR, apparent in the non-uniqueness of solutions to the
boundary value problem, vanishes in the discrete theory. However, in
the light of the analysis of \cite{OURS} and this article, one might
consider that not all of the configuration variables of Regge
calculus might be in fact physical. Rather, by constructing a
perfect action for discretized gravity, which reflects the continuum
dynamics and hence the gauge symmetries of GR, one might get more
insight into which of the degrees of freedom are actually physical,
and which are gauge. This is in particular important in attempts to
quantize discrete gravity theories by using Regge triangulations, as
happens in Spin Foams.

We therefore suggest that it might be valuable to study how gauge
degrees of freedom are regained in the continuum limit, and think
that the improved and perfect actions presented in this article can
be helpful in this pursuit.
\item In particular the usage of simplices with constant curvature
might be useful for first order formulations in Regge calculus, and
the questions of constraints in this context \cite{BIANCASYMM, OURS,
GP}. Furthermore, an area-angle formulation \cite{BDSS} with
simplices of constant (nonzero) curvature might be more viable than
in the flat case, since e.g. in $4D$ the $10$ dihedral angles of a
$4$-simplex determine its geometry completely, not just its
conformal structure as with flat simplices. These variables are not
only appropriate for spin foam models but seem also to be useful to
obtain a canonical formulation \cite{jr}. See \cite{newangles} for
formulations based on different sets of basic variables and a first
order formulation involving $\kappa$--curved simplices. Curved
simplices have been proposed in \cite{LF}, but no action has been
proposed there. In general quantum gravity models with a positive
cosmological constant are better behaved in the infrared and can
even serve as regulators for models without a cosmological constant.
Therefore it seems useful to investigate the construction of spin
foam models with a cosmological constant.

\item The Turaev--Viro invariant \cite{tuarevviro} for $3$-manifolds reproduces in the
semiclassical limit the geometry of constantly curved simplices for
$\Lambda>0$ \cite{taylor}. The construction of corresponding Spin
Foam models for $\Lambda<0$, which is still elusive, could benefit
form the formalism presented here by starting a quantization of the
perfect action (\ref{Gl:PerfectActionIn3D}) for $\Lambda<0$. In
general we note that for the $3D,\Lambda>0$ case a quantization
having the perfect action as a limit is available (namely the
Turaev--Viro models), whereas a similar quantization based on the
non--perfect action is missing. In the canonical formulation one has
to worry about complicated factor ordering ambiguities \cite{perez}
in addition to an anomalous constraint algebra. In contrast a
quantization based on $\kappa$--curved simplices could avoid these
issues.

In general it would be interesting to see whether a similar
procedure for reobtaining gauge symmetries (and triangulation
independence) as presented here for the classical theory works also
for the quantum theory.  The Ponzano--Regge with an added
cosmological term and the Tuarev--Viro model would be an interesting
example \cite{toappear2}. See also \cite{gp1} where spatial
diffeomorphism symmetry has been reobtained in the continuum limit
for a symmetry reduced model.
\end{itemize}

\section*{Acknowledgements} The authors would like to thank  John
Barrett, Simone Speziale, Ruth Williams and Jose Zapata for valuable
discussions and remarks. Funding of B.B. by ESF grant PESC/2805
within the \emph{Quantum Geometry and Quantum Gravity} network for a
visit in Utrecht is gratefully acknowledged. B.B. would like to
thank for the hospitality at the ITF, Utrecht. The research of B.D.
at the University of Utrecht was supported by a Marie-Curie
fellowship of the European Union.


\appendix

\section{Curved Simplices}\label{Ch:AppendixCurvedSimplices}

In the following, let $\sigma$ denote a $D$-dimensional simplex of
constant curvature $\kappa$. Denote its $D$-dimensional volume by
$V_\sigma^{(\kappa)}$. A hinge $h$ is a $D-2$ dimensional subsimplex
(which is again a simplex of constant curvature $\kappa$), and we
denote its $D-2$-dimensional volume by $F_h^{(\kappa)}$. For a hinge
$h\subset\sigma$ denote the interior deficit angle between the two
$D-1$-dimensional subsimplices of $\sigma$ meeting at $h$ by
$\theta_h^{(\kappa)}$.

The simplex $\sigma$ is completely determined by the lengths of its
$N:=\frac{D(D+1)}{2}$ edges (the $1$-simplices). All of the above
are regarded as functions of their lengths $L_1,\ldots,L_N$.\\

If we numerate the vertices of $\sigma$ from $1$ to $D+1$, we
specify a subsimplex by $(ij\ldots k)$ if it is spanned by the
vertices with the numbers $i,j\,\ldots, k$, and by $[ij\ldots k]$ if
it is spanned by all vertices \emph{except} $i,j,\ldots,k$. In this
notation an edge can be denoted as $e=(ij)$, and its dual hinge by
$h=[ij]$.

Denote the geodesic lengths of the edges $(ij)$ by $L_{(ij)}$. Then
the $(D+1)\times(D+1)$ matrix $G$ with entries
\begin{eqnarray}
G_{ij}\;=\;c_\kappa(L_{(ij)})
\end{eqnarray}

\noindent where the function $c_\kappa(x)$ is defined by
\begin{eqnarray*} \label{ckappa}
c_\kappa(x)\;:=\;\left\{\begin{array}{ll}
\cos\big(\sqrt\kappa x\big)&\quad \kappa>0\\[5pt]
\cosh\big(\sqrt{-\kappa} x\big)&\quad \kappa<0\end{array}\right.
\end{eqnarray*}

\noindent is called the \emph{Gram matrix} of the simplex. We denote
by $G^{ij}$ the inverse of $G_{ik}$.
Then the interior dihedral angle $\theta_{[ij]}$ opposite of the
edge $(ij)$ is given by \cite{KOKK1}
\begin{eqnarray}\label{dihedral}
\cos\theta_{[ij]}^{(\kappa)}\;=\;-\frac{G^{ij}}{\sqrt{G^{ii}}\sqrt{G^{jj}}}.
\end{eqnarray}

\noindent Hence, for any hinge $h$ the exterior angle
$\theta_h^{(\kappa)}$, regarded as a function of the lengths
$L_1,\ldots,L_N$, exhibits the scaling behavior\footnote{The
formulae presented here hold for $\kappa>0$. For $\kappa<0$
analogous formulae can be deduced.}
\begin{eqnarray}
\theta_h^{(\kappa)}(L_1,\ldots,L_N)\;=\;\theta_h^{(1)}(\sqrt\kappa
L_1,\ldots,\sqrt\kappa  L_N).
\end{eqnarray}

\noindent As a result we have
\begin{eqnarray}\label{Gl:DerivationOfAngle}
\frac{\partial}{\partial
\kappa}\theta^{(\kappa)}_h\;=\;\frac{1}{2\kappa}\sum_{e\subset
\sigma}L_e\frac{\partial \theta^{(\kappa)}_h}{\partial L_e}.
\end{eqnarray}

\noindent Furthermore, the geometric quantities in curved simplices
satisfy the Schl\"afli identity
\begin{eqnarray}\label{Gl:SchlaefliCurved}
\sum_{h\subset \sigma}F^{(\kappa)}_h\frac{\partial
\theta_h^{\sigma\,(\kappa)}}{\partial
L_e}\;=\;(D-1)\kappa\frac{\partial V_\sigma^{(\kappa)}}{\partial
L_e}\qquad\text{for all edges }e\subset\sigma
\end{eqnarray}

\subsection{Geometric Identities in curved
simplices}\label{Ch:AppendixGeometricIdentity}

In this section we derive a generalization of \emph{Euler's theorem}
\begin{eqnarray}\label{Gl:EulersTheorem}
\frac{1}{D}\sum_{e\subset\sigma}L_e\frac{\partial V_\sigma}{\partial
L_e}\;=\;V_\sigma
\end{eqnarray}

\noindent to curved tetrahedra.
\begin{Lemma} For a simplex $\sigma$ of dimension $D$ and constant
curvature we have
\begin{eqnarray}\label{Gl:ScalingOfVolume}
V^{(\kappa)}_\sigma(sL_1,\ldots, sL_N)\;=\;s^D\,V^{(\kappa
s^2)}_\sigma(L_1,\ldots, L_N)
\end{eqnarray}
\end{Lemma}

\noindent\textbf{Proof:} This can in fact be seen easily for
$\kappa>0$, where the simplex is a subset of a $D$-dimensional
sphere of radius $R=1/\sqrt{\kappa}$. If the radius is scaled by
$s$, as well as all the edge lengths, the volume of the sphere is
scaled by $s^D$, hence also the volume of the simplex. For
$\kappa<0$ a similar reasoning for hyperbolic spheres applies. The
formula (\ref{Gl:ScalingOfVolume}) follows. \qed

\begin{Corollary}
For any $D$-dimensional simplex $\sigma$ of constant curvature
$\kappa$ we have
\begin{eqnarray}\label{Gl:ScalingOfDerivative}
\frac{\partial }{\partial
s}V^{(\kappa)}_\sigma(L_1,\ldots,L_N)\;=\;s^{D-1}\sum_{e\subset\sigma}
L_e\frac{\partial }{\partial L_e}V^{(\kappa s^2)}_\sigma(L_1,\ldots,
L_N).
\end{eqnarray}
\end{Corollary}

\noindent\textbf{Proof:} By explicit calculation:
\begin{eqnarray*}
\frac{\partial }{\partial
s}V^{(\kappa)}_\sigma(sL_1,\ldots,sL_N)\;&=&\;\sum_{e\subset\sigma}
L_e\frac{\partial }{\partial (sL_e)}V^{(\kappa)}_\sigma(sL_1,\ldots,
sL_N)\\[5pt]
&=&\;\frac{1}{s}\sum_{e\subset\sigma} L_e\frac{\partial }{\partial
L_e}V^{(\kappa)}_\sigma(sL_1,\ldots, sL_N)\\[5pt]
&=&\;s^{D-1}\sum_{e\subset\sigma} L_e\frac{\partial }{\partial
L_e}V^{(\kappa s^2)}_\sigma(L_1,\ldots, L_N).
\end{eqnarray*}

\noindent This was the claim.\qed\\[5pt]

Another important identity is the following generalization of
Euler's formula to curved simplices:
\begin{Lemma}
If $\sigma$ is a $D$-dimensional simplex of constant curvature
$\kappa$, then
\begin{eqnarray}\label{Gl:MainFormula}
\frac{1}{D}\sum_{e\subset \sigma} L_e\frac{\partial
V^{(\kappa)}_\sigma}{\partial
L_e}\;=\;V_\sigma^{(\kappa)}\,+\,\frac{2}{D}\kappa\frac{\partial
V^{(\kappa)}_\sigma}{\partial \kappa}
\end{eqnarray}
\end{Lemma}

\noindent\textbf{Proof:} We prove this by induction over $D$, and
first note that it is trivially true for $D=1$. The case $D=2$ can
be shown explicitly by recalling the formula for the area of a
spherical (or hyperbolical) triangle $t$
\begin{eqnarray}\label{Gl:FormulaForAreaOfCurvedTriangle}
V_t^{(\kappa)}\;=\;\frac{\theta_1^{(\kappa)}+\theta_2^{(\kappa)}+\theta_3^{(\kappa)}-\pi}{\kappa}
\end{eqnarray}

\noindent where the $\theta_i^{(\kappa)}$ are the interior angles of
$t$. Since they are interior dihedral angles of curved simplices,
they satisfy the relations (\ref{Gl:DerivationOfAngle}). This leads
to
\begin{eqnarray}\label{Gl:DerivativeOfCurvedAreasWRTKappa}
\frac{\partial V_t^{(\kappa)}}{\partial
\kappa}\;=\;-\frac{V_t^{(\kappa)}}{\kappa}\;+\;\frac{1}{2\kappa}\sum_{i=1}^3L_i\frac{\partial
V_t^{(\kappa)}}{\partial L_i}.
\end{eqnarray}

\noindent This shows (\ref{Gl:MainFormula}) for $D=2$. We now show
that the formula is true for $D$ if it is true for $D-2$. We begin
with Schlaefli's formula \cite{SCHLENKER}
\begin{eqnarray}\label{Gl:Schlaefli}
(D-1)\kappa\,dV^{(\kappa)}_\sigma\;=\;\sum_{h\subset\sigma}
F_h^{(\kappa)}\,d\theta^{(\kappa)}_h
\end{eqnarray}

\noindent As a consequence, we have (whenever a function appears
without arguments, it is supposed to be taken at the point
$(L_1,\ldots, L_N)$):
\begin{eqnarray*}
(D-1)\kappa\;V_\sigma^{(\kappa)}\;&=&\;\int_0^1ds\,\sum_{h\subset\sigma}F_h^{(\kappa)}(sL_1,\ldots,sL_n)\,\frac{\partial
}{\partial s}\theta^{(\kappa)}_h(sL_1,\ldots,sL_N)\\[5pt]
&=&\;\sum_{h\subset
\sigma}F_h^{(\kappa)}\theta_h^{(\kappa)}\;-\;\int_0^1ds\sum_{h\subset
\sigma}\theta^{(\kappa s^2)}(L_1,\ldots,
L_N)\frac{\partial}{\partial s}F_h^{(\kappa)}(sL_1,\ldots sL_N)
\end{eqnarray*}

\noindent Remembering that each hinge $h$ is a $D-2$ dimensional
simplex of constant curvature $\kappa$, we conclude with
(\ref{Gl:ScalingOfDerivative}) that
\begin{eqnarray}\nonumber
S^{(\kappa)}\;:=\;-\sum_{h\subset
\sigma}F_h^{(\kappa)}\theta_h^{(\kappa)}\,+\,(D-1)\kappa
V^{(\kappa)}_\sigma\;&=&\;-\int_0^1ds\,s^{D-3}\sum_{h\subset
\sigma}\theta^{(\kappa s^2)}_h\sum_{e\subset h}L_e\frac{\partial
F_h^{(\kappa s^2)}}{\partial L_e}\\[5pt]\nonumber
&=&\;-\frac{1}{2}\kappa^{-\frac{D-2}{2}}\int_0^{\kappa}dy\;y^{\frac{D-4}{2}}\sum_{h\subset
\sigma}\theta_h^{(y)}\sum_{e\subset h}L_e\frac{\partial
F^{(y)}_h}{\partial L_e}\\[5pt]\label{Gl:CooleFormel}
\end{eqnarray}

\noindent where we have used a change of variable $y=\kappa s^2$.\\

We now derive the two different ways (\ref{Gl:CooleFormel}) of
writing $S^{(\kappa)}$ with respect to $\kappa$. The first one gives
us
\begin{eqnarray}\nonumber
\frac{\partial S^{(\kappa)}}{\partial
\kappa}\;&=&\;\frac{\partial}{\partial \kappa}\left(-\sum_{h\subset
\sigma}F_h^{(\kappa)}\theta_h^{(\kappa)}\,+\,(D-1)\kappa
V^{(\kappa)}_\sigma\right)\\[5pt]\label{Gl:DerivingSWRTKappaFirstForm}
&=&\;-\sum_{h\subset \sigma}\frac{F_h^{(\kappa)}}{\partial
\kappa}\theta_h^{(\kappa)}\,-\,\sum_{h\subset
\sigma}F_h^{(\kappa)}\frac{\partial \theta_h^{(\kappa)}}{\partial
\kappa}\,+\,(D-1)V_\sigma^{(\kappa)}\,+\,(D-1)\kappa\frac{\partial
V_\sigma^{(\kappa)}}{\partial \kappa}.
\end{eqnarray}

\noindent Note that with (\ref{Gl:DerivationOfAngle}) and
(\ref{Gl:Schlaefli}) we have

\begin{eqnarray}\nonumber
\sum_{h\subset \sigma}F_h^{(\kappa)}\frac{\partial
\theta_h^{(\kappa)}}{\partial
\kappa}\;&=&\;\frac{1}{2\kappa}\sum_{h\subset
\sigma}F_h^{(\kappa)}\sum_{e\subset \sigma}L_e\frac{\partial
V^{(\kappa)}_\sigma}{\partial L_e}\\[5pt]\label{Gl:UsefulFormula1}
&=&\;\frac{D-1}{2}\sum_{e\subset\sigma}L_e\frac{\partial
V_\sigma^{(\kappa)}}{\partial L_e}.
\end{eqnarray}

\noindent Now we use the induction hypothesis, which means that
(\ref{Gl:MainFormula}) in particular holds for $h$, i.e.

\begin{eqnarray}\label{Gl:UsefulFormula2}
\frac{\partial F_h^{(\kappa)}}{\partial
\kappa}\;=\;\frac{1}{2\kappa}\sum_{e\subset h}L_e\frac{\partial
F_h^{(\kappa)}}{\partial L_e}\,-\,\frac{D-2}{2\kappa}F_h^{(\kappa)}.
\end{eqnarray}

\noindent Inserting (\ref{Gl:UsefulFormula1}) and
(\ref{Gl:UsefulFormula2}) into (\ref{Gl:DerivingSWRTKappaFirstForm})
we arrive at

\begin{eqnarray}\nonumber
\frac{\partial S^{(\kappa)}}{\partial
\kappa}\;&=&\;-\frac{1}{2\kappa}\sum_{h\subset
\sigma}\theta_h^{(\kappa)}\sum_{e\subset h}L_e\frac{\partial
F_h^{(\kappa)}}{\partial L_e}\,+\,\frac{D-2}{2\kappa}\sum_{h\subset
\sigma}\theta_h^{(\kappa)}F_h^{(\kappa)}\\[5pt]\label{Gl:DerivingSWRTKappaFirstForm2}
&&\,-\,\frac{D-1}{2}\sum_{e\subset\sigma}L_e\frac{\partial
V_\sigma^{(\kappa)}}{\partial
L_e}\,+\,(D-1)V_\sigma^{(\kappa)}\,+\,(D-1)\kappa\frac{\partial
V_\sigma^{(\kappa)}}{\partial \kappa}.
\end{eqnarray}

\noindent On the other hand, by (\ref{Gl:CooleFormel}) we have

\begin{eqnarray}\nonumber
\frac{\partial S^{(\kappa)}}{\partial
\kappa}\;&=&\;\frac{\partial}{\partial
\kappa}\left(-\frac{1}{2}\kappa^{-\frac{D-2}{2}}\int_0^{\kappa}dy\;y^{\frac{D-4}{2}}\sum_{h\subset
\sigma}\theta_h^{(y)}\sum_{e\subset h}L_e\frac{\partial
F^{(y)}_h}{\partial
L_e}\right)\\[5pt]\label{Gl:DerivingSWRTKappaSecondForm}
&=&\;-\frac{D-2}{2\kappa}S^{(\kappa)}\;-\;\frac{1}{2\kappa}\sum_{h\subset
\sigma}\theta_h^{(\kappa)}\sum_{e\subset h}L_e\frac{\partial
F_h^{(\kappa)}}{\partial L_e}\\[5pt]\nonumber
&=&\;\frac{D-2}{2\kappa}\sum_{h\subset
\sigma}\theta_h^{(\kappa)}F_h^{(\kappa)}\,-\,\frac{(D-1)(D-2)}{2}V^{\kappa}_\sigma\;-\;\frac{1}{2\kappa}\sum_{h\subset
\sigma}\theta_h^{(\kappa)}\sum_{e\subset h}L_e\frac{\partial
F_h^{(\kappa)}}{\partial L_e}.
\end{eqnarray}

\noindent By comparing (\ref{Gl:DerivingSWRTKappaFirstForm2}) and
(\ref{Gl:DerivingSWRTKappaSecondForm}) we arrive at

\begin{eqnarray}
-\,\frac{D-1}{2}\sum_{e\subset\sigma}L_e\frac{\partial
V_\sigma^{(\kappa)}}{\partial
L_e}\,+\,(D-1)V_\sigma^{(\kappa)}\,+\,(D-1)\kappa\frac{\partial
V_\sigma^{(\kappa)}}{\partial
\kappa}\;=\;-\frac{(D-1)(D-2)}{2}V^{(\kappa)}_\sigma
\end{eqnarray}

\noindent which is equivalent to (\ref{Gl:MainFormula}).\qed

There is an important corollary: Deriving (\ref{Gl:MainFormula}) w.r.t $\kappa$ and setting $\kappa=0$, one can see that
\begin{eqnarray}
 (D+2)\frac{\partial V_\sigma^{(\kappa)}}{\partial \kappa}_{\Big|_{\kappa=0}}\;=\;\sum_{e\subset \sigma}l_e\frac{\partial }{\partial l_e}\frac{\partial V_\sigma^{(\kappa)}}{\partial \kappa}_{\Big|_{\kappa=0}},
\end{eqnarray}

\noindent which, by Euler's theorem, shows that $\frac{\partial
V_\sigma^{(\kappa)}}{\partial \kappa}_{|_{\kappa=0}}$ is a
homogenous function of the edge lengths $l_e$ of degree $D+2$. An
explicit example for this is e.g. $D=2$, where one can, with
(\ref{Gl:FormulaForAreaOfCurvedTriangle}), show that
\begin{eqnarray}
 \frac{\partial a_t^{(\kappa)}}{\partial \kappa}_{\Big|_{\kappa=0}}\;=\;\frac{1}{24}\sum_{e\subset t}l_e^2\;a_t^{(\kappa=0)},
\end{eqnarray}

\noindent which is indeed homogenous of degree $4$.

\end{document}